%% file: orderingNPHardness.tex
	\def\fullVersion{}
	\documentclass[11pt]{article}

\input{macros.tex}

\begin{document}

\title{On the \NP-Hardness of Approximating\\ Ordering Constraint
  Satisfaction Problems}%
\author{Per~Austrin \and Rajsekar~Manokaran \and Cenny~Wenner}

\maketitle  

\begin{abstract}
  We show improved $\NP$-hardness of approximating
  \emph{Ordering Constraint Satisfaction Problems}~(OCSPs).  For the
  two most well-studied OCSPs, \textproblem{Maximum Acyclic Subgraph}
  and \textproblem{Maximum Betweenness}, we prove
  inapproximability of
  $14/15+\epsilon$ and $1/2+\epsilon$.

  An OCSP is said to be approximation resistant if it is hard to
  approximate better than taking a uniformly random ordering.  We
  prove that the \textproblem{Maximum \emph{Non-}Betweenness Problem}
  is approximation resistant and that there are width-$m$
  approximation-resistant OCSPs accepting only a fraction $1 / (m/2)!$
  of assignments.
  These results provide the first
  examples of approximation-resistant OCSPs subject only to $\classP \neq \NP$.

\end{abstract}

\tableofcontents

\input{intro.tex}
\input{prelims.tex}
\input{general_hardness.tex}
\input{applications.tex}

\input{general_soundness_proof.tex}

\input{conclusion.tex}

\bibliographystyle{alpha}
\bibliography{ordering}




\end{document}

%% file: macros.tex
\input{preamble.tex}

\input{tcsMacros/stdinc.tex}

\input{tcsMacros/thmstyles.tex}

\ifx\includeOptSkipThmEnvironments\undefined\else
	\usepackage{thmtools}

	\ifdefined\theorem\else\declaretheorem[within=section]{theorem}\fi
	\ifdefined\corollary\else\declaretheorem[sibling=theorem]{corollary}\fi
	\ifdefined\example\else\declaretheorem[sibling=theorem]{example}\fi
	\ifdefined\lemma\else\declaretheorem[sibling=theorem]{lemma}\fi
	\ifdefined\proposition\else\declaretheorem[sibling=theorem]{proposition}\fi
	\ifdefined\claim\else\declaretheorem[sibling=theorem]{claim}\fi
	\ifdefined\definition\else\declaretheorem[sibling=theorem]{definition}\fi
	\ifdefined\remark\else\declaretheorem[sibling=theorem]{remark}\fi
	\ifdefined\conjecture\else\declaretheorem[sibling=theorem]{conjecture}\fi
        \spnewtheorem{procedure}{Procedure}{\bfseries}{\itshape}

\fi
\declaretheorem[sibling=theorem]{proposition}
\input{tcsMacros/prob-macros.tex}

\input{tcsMacros/mathy.tex}
\input{tcsMacros/optim-symbs.tex}
\input{tcsMacros/names.tex}

\input{tcsMacros/czoo.tex}

\newcommand{\ifSuper}[1][]{\ifblank{#1}{}{^{\Paren{#1}}}}

\newcommand{\full}[1]{\ifdefined\fullVersion {#1} \else {} \fi}
\newcommand{\lncs}[1]{\ifdefined\lncsVersion {#1} \else {} \fi}
\crefname{claim}{Claim}{Claims}

\newcommand*{\textproblem}[1]{\textsc{#1}}

\newcommand{\UG}{\ComplexityFont{UG}\xspace}

\newcommand{\from}{\ensuremath{\leftarrow}}
\newcommand{\irange}[1]{\Brac{#1}}

\newcommand*{\complementalt}[1]{ \overline{#1} }
\newcommand*{\comp}[1]{\complementalt{#1}}
\newcommand*{\suchthat}{~|~}
\newcommand*{\varsuchthat}{\,:\,}

\newcommand*{\iverson}[2][]{1_{#1}\left\{{#2}\right\}}
\DeclareDocumentCommand{\norm}{O{}O{}m}{{%
\left\vert\hspace{-0.2ex}\left\lvert {#3} \right\vert\hspace{-0.2ex}\right\vert_{#1}^{#2}%
}}
\renewcommand*{\abs}[1]{ { \left\lvert #1 \right\rvert } } 
\newcommand*{\floor}[1]{\left\lfloor#1\right\rfloor}

\DeclareDocumentCommand{\Inf}{O{l}og}{{%
\mathrm{Inf}\IfValueTF{#2}{^{({#2})}}{}_{{#1}}\IfNoValueTF{#3}{}{\left({#3}\right)}%
}}
\DeclareDocumentCommand{\Totinf}{og}{{%
\mathrm{TotInf}\IfValueTF{#1}{^{({#1})}}{}\IfNoValueTF{#2}{}{\left({#2}\right)}%
}}
\newcommand*{\CoinfTerm}{cross influence}
\DeclareDocumentCommand{\Coinf}{O{\pi}O{1-\gamma}m}{%
\mathrm{CrInf}^{(#2)}_{#1}\left({#3}\right)%
}

\newcommand*{\lifted}[2][]{{{#2}'}^{#1}}

\renewcommand*{\Pr}{\Psymb}
\renewcommand*{\prob}[2][]{\Pr_{#1}\paren{ #2 }}
\renewcommand*{\Prob}[2][]{\Pr_{#1}\Paren{ #2 }}
\newcommand*{\expectationSymbol}{\E}

\renewcommand*{\Ex}[2][]{\expectationSymbol_{#1}\Brac{#2}}

\newcommand*{\condEx}[3][]{\,\mathbf{E}_{#1}\hspace{-0.3ex}\Brac{#2 \,\vert\, #3}}

\newcommand*{\noiseop}{{\rmT\hspace{-0.25ex}}}

\newcommand*{\mD}{\mathcal{D}}

\newcommand*{\rmL}{\mathrm{L}}

\newcommand*{\rmP}{\mathrm{P}}

\newcommand*{\mT}{\mathcal{T}}
\newcommand*{\rmT}{\mathrm{T}}

\newcommand*\tuple[1]{\ensuremath{\vec{#1}}}
\newcommand*\perm[1]{\ensuremath{\tuple{#1}}}
\newcommand*\rv[1]{\ensuremath{\mathbf{#1}}}
\newcommand*\rvv[1]{\vec{\rv{#1}}}
\newcommand*\vrv[1]{\rvv{#1}}
\newcommand*\mat[1]{\ensuremath{{\MakeUppercase{#1}}}}

\newcommand*\defineShorthands[1]{%
	\expandafter\def\csname bf#1\endcsname{{\bf#1}}%
	\expandafter\def\csname rm#1\endcsname{{\rm#1}}%
	\expandafter\def\csname it#1\endcsname{{\it#1}}%
	\expandafter\def\csname itbf#1\endcsname{{\it\bf#1}}%
	\expandafter\def\csname bfit#1\endcsname{{\bf\it#1}}%
	\expandafter\def\csname cal#1\endcsname{{\cal#1}}%
	\expandafter\def\csname rv#1\endcsname{\rv{#1}}%
	\expandafter\def\csname v#1\endcsname{\vec{#1}}%
	\expandafter\def\csname rvv#1\endcsname{\rvv{#1 }}%
	\expandafter\def\csname vrv#1\endcsname{\vrv{#1}}%
	\expandafter\def\csname mat#1\endcsname{\mat{#1}}%
	\expandafter\def\csname rvmat#1\endcsname{\rv{\mat{#1}}}%
	\expandafter\def\csname matrv#1\endcsname{\rv{\mat{#1}}}%
}
\defineShorthands{x}
\defineShorthands{y}
\defineShorthands{z}
\defineShorthands{X}
\defineShorthands{Y}
\defineShorthands{Z}

\newcommand*{\x}[1][]{\rvv{x}\ifSuper[#1]}
\newcommand*{\y}[1][]{\rvv{y}\ifSuper[#1]}
\newcommand*{\z}[1][]{\rvv{z}\ifSuper[#1]}

\newcommand*{\X}{\rv{\mat{X}}}
\newcommand*{\Y}{\rv{\mat{Y}}}

\newcommand{\Ctodo}[1]{}

\newcommand*{\payoff}{\mathcal{P}}

\newcommand*{\bndI}{[0, 1]}

\newcommand*{\pcomp}[1]{\bar{#1}}

\newcommand*{\lc}{\cL\xspace}
\newcommand*{\lcLeft}{U}
\newcommand*{\lcRight}{V}
\newcommand*{\lcEdges}{E}
\newcommand*{\lcLeftLabelSet}{L}
\newcommand*{\lcRightLabelSet}{R}
\newcommand*{\lcProj}{\Pi}

\newcommand*{\lcDesc}{\lc = (\lcLeft, \lcRight, \lcEdges,
  \lcLeftLabelSet, \lcRightLabelSet, \lcProj)\xspace}

\newcommand*{\labelcover}{%
\renewcommand*{\labelcover}{LC\xspace}\textproblem{Label Cover}~(LC)\xspace%
}

\newcommand*{\baseDist}{\mD}

\newcommand*{\decoup}[1]{{#1}^{\bot}}

\newcommand*{\ntestDist}[2][\pi]{\super{\mT_{#1}}{\gamma}\left(#2\right)}

\newcommand*{\ordr}{\cO}
\newcommand*{\ordrC}[1][c]{\ordr_{|#1}}
\newcommand*{\ocsp}{\text{OCSP}}

\newcommand*{\bucketPred}[1][]{\wp\ifSuper[#1]}

\newcommand*{\bucketInd}[1][a]{{F}\ifSuper[#1]}
\newcommand*{\bucketIndR}[1][a]{{G}\ifSuper[#1]}

\newcommand*{\dictDist}[1][]{\dict\ifSuper[#1]}

\newcommand{\instance}{\ensuremath\cI}
\newcommand{\outputV}{\cX}
\newcommand{\outputE}{\cC}

\newcommand*{\MAS}{\textproblem{MAS}\xspace}
\newcommand*{\MaxBTW}{\textproblem{Max BTW}\xspace}
\newcommand*{\MaxNBTW}{\textproblem{Max NBTW}\xspace}
\newcommand*{\MaxSO}{\textproblem{Max $2t$-SO}\xspace}

\newcommand*{\BTW}{\ensuremath{\mathrm{BTW}}\xspace}
\newcommand*{\NBTW}{\ensuremath{\mathrm{NBTW}}\xspace}
\newcommand*{\SOPred}[1][2t]{\ensuremath{{#1}\mathrm{\text{-}SO}}\xspace} 

\newcommand*\accP[1][f,g]{\mathsf{Acc}_{#1}}
\newcommand*\bVal[1][f,g]{\mathsf{BAcc}_{#1}}

\newcommand{\dict}{\cT}

\ifdefined\lncsVersion
\newcommand*\qedendproof{ \qed \end{proof}}
\else
\newcommand*\qedendproof{\end{proof}}
\fi

\newcommand*\eg{e.g.\@}
\newcommand*\ie{i.e.\@}

\def\gobbleminus#1{\ifx-#1\else#1\fi}
\def\IsInteger#1{%
  TT\fi
  \ifcat_\ifnum9<1\gobbleminus#1 _\else A\fi
}

\newcommand*\ordinal[1]{\if\IsInteger{#1}\engordnumber{#1}\else #1\engordletters{th}\fi}


%% file: preamble.tex
\usepackage{etoolbox, xparse}
\usepackage{xspace}
\usepackage{soul} 
\usepackage{boxedminipage}
\usepackage{tkz-graph}
\usepackage[font=small,labelfont=bf]{caption}
\usepackage{float}
\usepackage{wrapfig}

\usepackage{verbatim}
\usepackage{color,graphicx}
\usepackage{tabularx}
\usepackage{amsmath,amssymb,amsfonts}
\usepackage{fullpage}
\usepackage{nicefrac}
\usepackage{booktabs}
\usepackage{engord}

\usepackage{bm}

\ifx\includeOptExtraMath\undefined
\else
\usepackage[showonlyrefs]{mathtools}
\usepackage{mathstyle}
\usepackage{breqn}
\usepackage{empheq}
\fi

\usepackage{nameref}
\usepackage[colorlinks,linkcolor=blue,filecolor = blue,
citecolor = blue, urlcolor = blue]{hyperref}
\usepackage[capitalize]{cleveref}

\usepackage{textcomp,setspace}

%% file: tcsMacros/stdinc.tex
\newcommand{\bits}{\{0,1\}}

\newcommand{\defeq}{\stackrel{\mathrm{def}}=}
\newcommand{\supp}{\mathrm{supp}}
\newcommand{\set}[1]{\{#1\}}

\newcommand{\paren}[1]{(#1 )}
\newcommand{\Paren}[1]{\left(#1 \right )}

\newcommand{\Brac}[1]{\left[#1 \right]}
\newcommand{\abs}[1]{\lvert#1\rvert}

\newcommand{\card}[1]{\lvert#1\rvert}

\newcommand{\norm}[1]{\lVert#1\rVert}

\renewcommand{\vec}[1]{{\bm{#1}}}

\newcommand{\super}[2]{#1^{\paren{#2}}}

\definecolor{DSred}{rgb}{1,0,0}

\renewcommand{\epsilon}{\varepsilon}



%% file: tcsMacros/thmstyles.tex
\ifx\includeOptSkipThmEnvironments\undefined
\usepackage{amsthm}
\usepackage{thmtools}

\declaretheorem[within=section]{theorem}
\declaretheorem[sibling=theorem]{corollary}
\declaretheorem[sibling=theorem]{example}
\declaretheorem[sibling=theorem]{lemma}
\declaretheorem[sibling=theorem]{claim}

\declaretheorem[sibling=theorem]{definition}
\declaretheorem[sibling=theorem]{remark}
\declaretheorem[sibling=theorem]{conjecture}
\declaretheorem[sibling=theorem]{procedure}

\fi

\newcounter{termcounter}
\renewcommand{\thetermcounter}{\Alph{termcounter}}
\crefname{term}{term}{terms}
\creflabelformat{term}{#2\textup{(#1)}#3}

\makeatletter
\def\term{\@ifnextchar[\term@optarg\term@noarg}
\def\term@optarg[#1]#2{%
  \textup{#1}%
  \def\@currentlabel{#1}%
  \def\cref@currentlabel{[][2147483647][]#1}%
  \cref@label[term]{#2}}
\def\term@noarg#1{%
  \refstepcounter{termcounter}%
  \textup{\thetermcounter}%
  \cref@label[term]{#1}}
\makeatother

%% file: tcsMacros/prob-macros.tex
\newcommand{\Esymb}{{\bf E}}

\newcommand{\Psymb}{{\bf P}}

\DeclareMathOperator*{\E}{\Esymb}

\renewcommand{\Pr}{\ProbOp}

\newcommand{\prob}[1]{\Pr\set{ #1 }}
\newcommand{\Prob}[1]{\Pr\Set{ #1 }}

\newcommand{\Ex}[1]{\E\Brac{#1}}



%% file: tcsMacros/mathy.tex
\newcommand{\R}{\mathbb{R}}

\newcommand{\Z}{\mathbb{Z}}

\newcommand{\cC}{\mathcal C}
\newcommand{\cD}{\mathcal D}

\newcommand{\cI}{\mathcal I}

\newcommand{\cL}{\mathcal L}

\newcommand{\cO}{\mathcal O}
\newcommand{\cP}{\mathcal P}

\newcommand{\cR}{\mathcal R}

\newcommand{\cT}{\mathcal T}

\newcommand{\cX}{\mathcal X}


%% file: tcsMacros/optim-symbs.tex
\newcommand{\val}{{\sf val}}

\newcommand{\Inf}{\mathrm{Inf}}



%% file: tcsMacros/names.tex
\newcommand{\Hastad}{H\aa{}stad\xspace}

\newcommand{\JHastad}{Johan~\Hastad}

%% file: tcsMacros/czoo.tex
\newcommand{\ComplexityFont}[1]{\ensuremath{\mathsf{#1}}}

\newcommand{\classP}{\ComplexityFont{P}}

\newcommand{\NP}{\ComplexityFont{NP}}

%% file: intro.tex


\section{Introduction}




We study the $\NP$-hardness of approximating a rich class of
optimization problems known as the \emph{Ordering Constraint
  Satisfaction Problems}~(OCSPs).  An instance of an OCSP is described
by a set of variables $\outputV$ and a set of \emph{local ordering
  constraints} $\outputE$.  Each constraint specifies a set of
variables and a set of permitted permutations of these variables.  The
objective is to find a permutation of $\outputV$ that maximizes the
fraction of constraints satisfied by the induced local permutations.

A simple example of an OCSP is the \textproblem{Maximum Acyclic
  Subgraph}~(\MAS) where one is given a directed graph $G =
(V, A)$ with the task of finding an acyclic subgraph of $G$ with the
maximum number of edges.  Phrased as an OCSP, $V$ is the set of
variables and each edge $u \to v$ is a constraint ``$u \prec v$''
dictating that $u$ should precede $v$.  The maximum fraction of
constraints simultaneously satisfiable by an ordering of $V$ is then
exactly the normalized size of the largest acyclic subgraph,
also called the value of the
instance.  Since the constraints in an \MAS instance are on two
variables, it is an OCSP of width two. Another example of an OCSP is
the \textproblem{Maximum Betweenness}~(\MaxBTW) problem ~\cite{GJ79}.
In this width-three OCSP, a constraint on a triplet of variables $(x,y,z)$
is satisfied by the local ordering $x \prec z \prec y$
and its reverse, $y \prec z \prec x$; in other words, $z$ has to be
between $x$ and $y$, giving rise to the name for the problem.

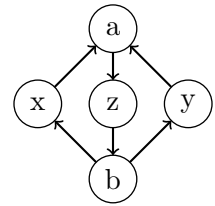
\begin{wrapfigure}{r}{3.4cm}
\label{fig:masgadget}
\label{fig:mas_section_mas_gadget}
\begin{center}
\begin{tikzpicture}[->,node distance=1.05cm]
\tikzset{EdgeStyle/.style={->,}}
\Vertex{x}
\EA(x){z}
\EA(z){y}
\NO(z){a}
\SO(z){b}
\Edges(b,x,a,z,b,y,a)
\end{tikzpicture}
\end{center}
\vspace{-0.5ex}
\caption[]{An \MAS instance with value $5/6$.}
\end{wrapfigure}

Determining the value of a \MAS instance is already $\NP$-hard and one
turns to approximation algorithms. An algorithm is called a $c$-approximation
if, when applied to an instance $\instance$, it is guaranteed to produce
an ordering satisfying at least a fraction $c \cdot \val(\instance)$ of
the constraints.
Every OCSP
admits a naive approximation algorithm which picks an ordering
of $\outputV$ uniformly at random without even looking at the
constraints.  For \MAS, this algorithm yields a $1/2$-approximation in
expectation as each constraint is satisfied with probability $1/2$
for a random ordering.
Surprisingly, there is evidence that this mindless procedure
achieves the best approximation ratio possible in polynomial time:
assuming Khot's Unique Games Conjecture~(UGC)~\cite{Khot02a}, \MAS is
hard to approximate within $1/2+\epsilon$ for every $\epsilon >
0$~\cite{GuruswamiMR08,GuruswamiHMRC11}.  An OCSP is called
\emph{approximation resistant} if it exhibits this behavior, i.e., if
it is \NP-hard to improve on the guarantee of the random-ordering
algorithm by $\epsilon$ for every $\epsilon > 0$.  In fact, the
results of \cite{GuruswamiHMRC11} are much more general: assuming the
UGC, they prove that \emph{every} OCSP of bounded width is
approximation resistant.

\Ctodo{This paragraph doesn't flow naturally in my opinion}
In many cases -- such as for \textproblem{Vertex Cover},
\textproblem{Max Cut}, and as we just mentioned, for all OCSPs --
the UGC allows us to prove optimal $\NP$-hard inapproximability results
which are not known without the conjecture.  For instance, the
problems \MAS and \MaxBTW were to date only known to be $\NP$-hard to
approximate within $65/66+\epsilon$~\cite{Newman01} and
$47/48+\epsilon$~\cite{ChorS98}, which comes far from matching the
random assignment thresholds of $1/2$ and $1/3$, respectively.  In
fact, while the UGC implies that all OCSPs are approximation
resistant, there were no results proving $\NP$-hard approximation
resistance of an OCSP prior to this work.  In contrast, there is a
significant body of work on \NP-hard approximation resistance of
classical Constraint Satisfaction
Problems~(CSPs)~\cite{Hastad01,SamorodnitskyT00,EngebretsenH08,Chan13}.
Furthermore, the UGC is still very much open and recent algorithmic
advances have given rise to subexponential algorithms for Unique
Games~\cite{AroraBS10,BarakBHKSZ12} putting the conjecture in
question.  Several recent works have also been aimed at bypassing the
UGC for natural problems by providing comparable results without
assuming the conjecture~\cite{GuruswamiRSW12,Chan13}.



\subsection{Results}\label{sec:results}%

\begin{table}
\begin{center}
\newcommand*\supcite[1]{ {$^{\textrm{\cite{#1}} } $} }%
\begin{tabular}{llllll}
Problem &%
Approx. factor &%
\UG-inapprox. &%
\NP-inapprox.&%
This work\\%
\hline%
\MAS &%
$1/2 + \Omega(\nicefrac{1}{\log{n}})$\supcite{CharikarMM07}&%
$1/2 + \epsilon$\supcite{GuruswamiMR08} &%
$65/66 + \epsilon$\supcite{Newman01} &%
$14/15 + \epsilon$%
\\%
\MaxBTW &%
$1/3$ &%
$1/3 + \epsilon$\supcite{CharikarGM09} &%
$47/48 + \epsilon$\supcite{ChorS98} &%
$1/2 + \epsilon$%
\\%
\MaxNBTW &%
$2/3$ &%
$2/3 + \epsilon$\supcite{CharikarGM09} &%
- &%
$2/3 + \epsilon$%
\\%
$m$-OCSP &%
$1/m!$ &%
$1/m! + \epsilon$\supcite{GuruswamiHMRC11}  &%
- &%
$1 / \floor{m / 2}! + \epsilon$%
\end{tabular}%
\end{center}%
\caption{Known results and our improvements.}%
\label{table:ocsp_results}%
\end{table}




In this work we obtain improved $\NP$-hardness of approximating
various OCSPs.  While a complete characterization such as in the UG
regime still eludes us, our results improve the knowledge of what we
believe are four important flavors of OCSPs; see
\cref{table:ocsp_results} for a summary of the present state of
affairs.

We address the two most studied OCSPs: \MAS and \MaxBTW. For \MAS,
we show a factor $(14/15+\epsilon)$-inapproximability improving the factor from
$65/66+\epsilon$~\cite{Newman01}.  For \MaxBTW, we show a factor
$(1/2+\epsilon)$-inapproximability improving from $47/48+\epsilon$~\cite{ChorS98}.

\begin{theorem}\label{thm:mas_result}%
  For every $\epsilon > 0$, it is $\NP$-hard to distinguish between
  \MAS instances with value at least $15/18-\epsilon$ from instances
  with value at most $14/18 + \epsilon$.
\end{theorem}

\begin{theorem}\label{thm:mbtw_result}%
  For every $\epsilon > 0$, it is $\NP$-hard to distinguish between
  \MaxBTW instances with value at least $1-\epsilon$ from instances with
  value at most $1/2 + \epsilon$.
\end{theorem}

The above two results are inferior to what is known assuming the UGC
and in particular do not prove approximation resistance.  We introduce
the \textproblem{Maximum Non-Betweenness}~(\MaxNBTW) problem which
accepts the \emph{complement} of the predicate in \MaxBTW.  This
predicate accepts four of the six permutations on three
elements and thus a random ordering satisfies two thirds of the constraints
in expectation.  We show that this is optimal up to smaller-order
terms.

\begin{theorem}\label{thm:mnbtw_result}%
  For every $\epsilon > 0$, it is $\NP$-hard to distinguish between
  \MaxNBTW instances with value at least $1-\epsilon$ from instances with
  value at most $2/3 + \epsilon$.
\end{theorem}

Finally, we address the approximability of a generic width-$m$ OCSP as
a function of the width $m$.  In the CSP world, the generic
version is called $m$-CSP and we call the ordering version $m$-OCSP.
We devise a simple predicate, ``$2t$-Same Order'' (\SOPred),
on $m=2t$ variables that is satisfied only if the first
$t$ elements are relatively ordered exactly as the last $t$ elements.
A random ordering satisfies only a fraction $1/t!$ of the constraints and we
prove that this is essentially optimal, implying a
$(1/\floor{m/2}!+\epsilon)$-factor inapproximability of $m$-OCSP.

\begin{theorem}\label{thm:mocsp_result}%
  For every $\epsilon > 0$ and integer $m \geq 2$, it is $\NP$-hard to
  distinguish $m$-\textrm{OCSP} instances with value at least $1 - \epsilon$
  from value at most $1 / \floor{m / 2}! + \epsilon$.
\end{theorem}

\subsection{Proof Overview}
With the exception of \MAS, our results follow a route which is by now
standard in inapproximability: starting from the optimization problem
\labelcover, we give a reduction to the problem at hand using a
\emph{dictatorship-test} gadget, also known as a long-code test.
We describe this reduction in the context of
\MaxNBTW to highlight the new techniques in this paper.

The reduction produces an instance $\instance$ of \MaxNBTW from an
instance $\lc$ of \labelcover such that $\val(\instance)
> 1 - \eta$ if $\val(\lc) = 1$ while $\val(\instance) < 2/3 + \eta$ if
$\val(\lc) \le \delta$.  By the PCP Theorem and the Parallel
Repetition Theorem~\cite{AroraSafra98,AroraLMSS98,Raz98}, it is
$\NP$-hard to distinguish between $\val(\lc) = 1$ and $\val(\lc) \le
\delta$ for every constant $\delta > 0$ and thus we obtain the result
in \cref{thm:mnbtw_result}.
The core component in this paradigm is the design of a
dictatorship test: a \MaxNBTW instance on $[q]^\lcLeftLabelSet \cup [q]^\lcRightLabelSet$,
for integers $q$ and label sets $\lcLeftLabelSet$ and $\lcRightLabelSet$.
Let $\pi$ be a map $\lcRightLabelSet \to \lcLeftLabelSet$. Each
constraint is a tuple $(\x, \y, \z)$ where $\x \in [q]^L$, while $\y,
\z \in [q]^R$. The distribution of tuples is obtained as follows.
First, pick $\x$, and $\y$ uniformly at random from $[q]^L$, and
$[q]^R$.  Set $z_j = y_j + x_{\pi(j)} \bmod q$.  Finally, add noise by
independently replacing each coordinate $x_i$, $y_j$ and $z_j$ with a
uniformly random element from $[q]$ with probability $\gamma$.

This test instance has canonical assignments that satisfy almost all
the constraints.  These are obtained by picking an arbitrary $j \in
[R]$, and partitioning the variables into $q$ sets $S_0, \ldots
S_{q-1}$ where $S_t = \set{\x \in [q]^\lcRightLabelSet |\ x_{\pi(j)} = t} \cup
\set{\y \in [q]^\lcLeftLabelSet |\ y_j = t}.$ If a constraint $(\x, \y, \z)$ is
so that $\x \in S_t$, $\y \in S_u$ then $\z \notin S_v$ for any $v \in
\set{t+1, \ldots, u-1}$ except with probability $O(\gamma)$.  This is
because $(a+b)$ mod $q$ is never strictly between $a$ and $b$.
Further, the probability that any two of $\x$, $\y$, and $\z$ fall in
the same set $S_i$ is simply the probability that any two of
$x_{\pi(j)}, y_j$, and $z_j$ are assigned the same value,
which is at most $O(1/q)$.  Thus, ordering the variables such
that $S_0 \prec S_1 \prec \ldots \prec S_{q-1}$ with an arbitrary
ordering of the variables within a set satisfies a fraction $1 - O(1/q)
- O(\gamma)$ constraints.

The proof of \cref{thm:mnbtw_result} requires a partial converse of
the above: every ordering that satisfies more than a fraction $2/3 +
\epsilon$ of the constraints is more-or-less an ordering that depends
on a few coordinates $j$ as above.  This proof involves three steps.
First, we show that there is a $\Gamma = \Gamma(q, \gamma, \beta)$ such that
every ordering $\ordr$ of $\irange{q}^\lcLeftLabelSet$ or $\irange{q}^\lcRightLabelSet$
can be broken into $\Gamma$ sets $S_0, \ldots, S_{\Gamma-1}$
such that one achieves expected value
at least $\val(\ordr) - \beta$ for arbitrarily small $\beta$
by ordering the sets $S_0 \prec \ldots \prec S_{\Gamma-1}$
and within each set ordering elements randomly.
The proof of this ``bucketing''
uses hypercontractivity of noised functions from a finite
domain.  We note that a related bucketing argument is
used in proving inapproximability of OCSPs assuming the
UGC~\cite{GuruswamiMR08,GuruswamiHMRC11}.  Their bucketing argument
is tied to the use of the UGC, where $\card{L} = \card{R}$ for the corresponding
dictatorship test, and does not extend to our setting.  In particular,
our approach yields $\Gamma \gg q$ while they crucially require $\Gamma \ll
q$ in their work.  We believe that our bucketing argument is more general
and a useful primitive.

Then, similarly to \cite{GuruswamiMR08,GuruswamiHMRC11}, the bucketing
argument allows an OCSP to be analyzed as if it were a CSP on a finite
domain, enabling us to use powerful techniques developed in this
setting.  In particular, we show that unless $\val(\lc) > \delta$, the
distribution of constraints $(\x, \y, \z)$ can be regarded as obtained
by sampling $\x$ independently upto an error $\eta$ in the payoff; in
other words, $\x$ is ``decoupled'' from $(\y, \z)$.  We note that the
marginal distribution of the tuple $(\y, \z)$ is already symmetric
with respect to swaps: $\Prob{\y = y, \z = z} = \Prob{\y = z, \z =
  y}$.  In order to prove approximation resistance, we combine three
of these dictatorship tests: the \ordinal{$j$} variant has $\x$ as the
\ordinal{$j$}
component of the $3$-tuple.  We show that the combined instance is
symmetric with respect to every swap up to an error $O(\eta)$ unless
$\val(\lc) > \delta$.  This implies that the instance has value at
most $2/3 + O(\eta)$ hence proving approximation resistance of
\MaxNBTW.

For \MaxBTW and \MaxSO, we do not require the final symmetrization and
instead use a dictatorship test based on a different distribution.
Finally, the reduction to \MAS is a simple gadget reduction from
\MaxNBTW.  For hardness results of width-two predicates, such gadget
reductions presently dominate the scene of classical CSPs and also
define the state of affairs for \MAS.  As an example, the best-known
$\NP$-hard approximation hardness of $16 / 17 + \epsilon$ for
\textproblem{Max Cut}\ is via a gadget reduction from \textproblem{Max
  3-Lin-2}~\cite{Hastad01,TrevisanSSW00}.  The previously best
approximation hardness of \MAS was also via a gadget reduction from
\textproblem{Max 3-Lin-2}~\cite{Newman01}, although with the
significantly smaller gap $65/66 + \epsilon$. By reducing from a
problem more similar to \MAS{}, namely \MaxNBTW{}, we improve to the
approximation hardness to $14/15 + \epsilon$.  The gadget in question
is quite simple and we have in fact already seen it in
\cref{fig:mas_section_mas_gadget}.

\paragraph{Organization.}  \cref{sec:preliminaries} sets up the
notation used in the rest of the article. \cref{sec:dictatorshipTest}
gives a general hardness result based on a test distribution which is
subsequently used in \cref{sec:inapproxRedn} to derive our main results.  The
proof of the soundness of the general hardness reduction is largely given
in \cref{sec:proof of soundness}.


%% file: prelims.tex
\section{Preliminaries}\label{sec:preliminaries}

We denote by $[n]$ the integer interval $\{0,\dotsc,n-1\}$.
Given a tuple of reals $\vx \in \R^m$, we write $\perm\sigma(\vx) \in S_m$
for the natural-order permutation on $\{1, \ldots, m\}$ induced by $\vx$.
For a distribution $\baseDist$ over
$\Omega_1 \times \dotsm \times \Omega_m$, we use
$\baseDist_{\le t}$ and $\baseDist_{> t}$ to denote the projection to
coordinates up to $t$ and the remaining, respectively.

\subsection{Ordering Constraint Satisfaction Problems.}\label{sec:ocsp
  prelims}  We are concerned with predicates $\payoff: S_m \to [0,1]$ on the symmetric group
$S_m$.  Such a predicate specifies a width-$m$ OCSP written
as $\ocsp(\payoff).$ 
An instance $\instance$ of $\ocsp(\payoff)$ problem is a tuple
$(\outputV, \outputE)$ where $\outputV$ is the set of \emph{variables} and
$\outputE$ is a distribution over ordered $m$-tuples of $\outputV$
referred to as the \emph{constraints}.

An assignment to $\instance$ is an injective map $\ordr: \outputV \to
\Z$ called an ordering of $\outputV$.  For a tuple $\tuple{c} = (v_1, \ldots,
v_m)$, $\ordrC$ denotes the tuple $(\ordr(v_1), \ldots, \ordr(v_m))$.
An ordering is said to satisfy the constraint $c$ when $\payoff(\perm\sigma(\ordrC)) = 1$.
The value of an ordering is the probability that a
random constraint $c \from \outputE$ is satisfied by $\ordr$
and the value $\val(\instance)$ of an instance is the maximum
value of any ordering. Thus,
\[ \val(\instance) = \max_{\ordr: \outputV \to \Z} \val(\ordr; \instance)
= \max_{\ordr: \outputV \to \Z}\, \Ex[c \in \outputE] { \payoff(\ordrC) }. \]
We extend the definition of value to include orderings that are not
strictly injective as follows.  Extend the predicate to $\payoff: \Z^m
\to \bndI$ by setting $\payoff(a_1,\ldots,a_m) = \Ex[\perm\sigma] {
  \payoff(\perm\sigma) }$ where $\perm\sigma$ is drawn uniformly
  at random over all permutations in $S_m$ such that $\sigma_i < \sigma_j$ whenever
$a_i < a_j.$ Note that the value of an instance is unchanged by
this extension as there is always a complete ordering
that attains the value of a non-injective map.

We define the predicates and problems studied in this work.
\MAS is exactly
$\ocsp(\{(1,2)\})$.  The betweenness predicate \BTW is the set
$\{(1,3,2), (3,1,2)\}$ and $\NBTW$ is $S_3 \setminus \BTW$. We define
\MaxBTW as $\ocsp(\BTW)$ and \MaxNBTW as $\ocsp(\NBTW)$. Finally, introduce 
$\SOPred[2t]$ as the subset of $S_{2t}$ such that the
induced ordering on the first $t$ elements coincides with the ordering of
the last $t$ elements, \ie
\[ \SOPred[2t] \defeq \{ \pi \in S_{2t} \suchthat \sigma( \pi(1),
\dotsc, \pi(r) ) = \sigma( \pi(r+1), \dotsc, \pi(2t)
) \}. \]
This predicate has $\nicefrac{(2t)!}{t!}$ satisfying orderings and will be
used in proving the inapproximability of the generic $m$-OCSP
with constraints of width at most $m$.


\subsection{Label Cover and Inapproximability.}\label{sec:prel:lc}





The problem \labelcover is a common starting point of
strong inapproximability results. An \labelcover instance $\lcDesc$
consists of a bipartite graph $(U \cup V, E)$ associating with every edge
$u,v$ a projection $\pi_{uv} : \lcRightLabelSet \rightarrow \lcLeftLabelSet$ with the goal of
labeling the vertices, $\lambda : U \cup V \rightarrow \lcLeftLabelSet \cup \lcRightLabelSet$, to
maximize the fraction of projections for which `$\pi_{uv}(\lambda(v)) =
\lambda(u)$'.
The following well-known hardness result follows from the PCP Theorem~\cite{AroraLMSS98} and the Parallel Repetition Theorem~\cite{Raz98}.
\begin{theorem}\label{cthm:lcHardness}%
  For every $\epsilon > 0$, there exists fixed label sets $\lcLeftLabelSet$ and $\lcRightLabelSet$
  such that it is $\NP$-hard to distinguish \labelcover
  instances of value one from instances of value at most $\epsilon.$
\end{theorem}



\subsection{Primer on Real Analysis}\label{sec:realAnalysis}

\Ctodo{We are not consistently using :th notation. At places superscript, at places :th}
We refer to a finite domain $\Omega$ along with a distribution $\mu$
as a probability space.  Given a probability space $(\Omega, \mu)$,
the $\ordinal{n}$ tensor power is $(\Omega^n, \mu^{\otimes n})$
where $\mu^{\otimes n}(\omega_1, \dotsc, \omega_n) = \mu(\omega_1) \dotsm
\mu(\omega_n)$.  The $\ell_p$ norm of $f: \Omega \to \R$ w.r.t.~$\mu$
is denoted by $\norm[\mu,p]{f}$ and is defined as $\Ex[\rv{x} \sim
\mu]{\abs{f(\rv{x})}^p}^{1/p}$ for real $p \geq 1$ and $\max_{x \in
  \supp(\mu)} f(x)$ for $p = \infty$. When clear from the context, we
omit the distribution $\mu$.  The following noise operator and
its properties play a pivotal role in our analysis.

\begin{definition}
  Let $(\Omega, \mu)$ be a probability space and $f: \Omega^n
  \rightarrow \R$ be a function on the $\ordinal{n}$ tensor
  power. For a parameter $\rho \in [0,1]$, the \emph{noise operator}
  $\noiseop_{\rho}$ takes $f$ to $\noiseop_{\rho}f \rightarrow \R$
  defined by
  \[
  \noiseop_{\rho}f(\vx) = \Ex{f(\rvvy) | \vx},
  \]
  where the \ordinal{$i$} coordinate of $\rvvy$ is chosen as $\rvvy_i = \vx_i$ with
  probability $\rho$ and otherwise as an independent new sample.
\end{definition}

The noise operator preserves the mass $\Ex{f}$ of a
function while spreading it in the
space.  The quantitative bound on this is referred to as
hypercontractivity.

\begin{theorem}[\cite{ourMainManWolff}; Theorem $3.16$,
  $3.17$ of \cite{Mossel10}]%
\label{prop:awesomeHCEstimate}%
Let $(\Omega, \mu)$ be a probability space in which the
minimum nonzero probability of any atom is $\alpha < 1/2.$ Then, for
every $q > 2$ and every $f: \Omega^n \to \R$,
\[ \norm{\noiseop_{{\rho(q)}} f}_q \le \norm{f}_2, \] where for $\alpha < 1/2$ we set
$A = \frac{1-\alpha}{\alpha}$; $1 / q' = 1 - 1/q$; and
${\rho(q,\alpha)} = \left(\frac{A^{1/q} - A^{-1/q}}{
A^{1/q'} - A^{-1/q'}}\right)^{1/2}$.
For $\alpha = 1/2$, we set ${\rho(q)} = (q-1)^{-1/2}$.
\end{theorem}

For a fixed probability space, the above theorem says that for
every $\gamma > 0$, there is a $q > 2$ such that
$\norm{\noiseop_{1-\gamma} f}_q \le \norm{f}_2$. For our application,
we need the easy corollary that the reverse
direction also holds: for every $\gamma > 0$, there exists a $q > 2$
such that hypercontractivity to the $\ell_2$-norm holds.

\Ctodo{Perhaps someone should verify this, or we just defer it to the
full version as it is a believable corollary.}
\begin{lemma}\label{prop:bestHCBound}%
  Let $(\Omega, \mu)$ be a probability space in which the minimum
  nonzero probability of any atom is $\alpha \le 1/2$. Then, for
  every $f: \Omega^n \to \R$, small enough $\gamma > 0$,
  \[ \norm{\noiseop_{1-\gamma} f}_{2+\delta} \le \norm{f}_2 \] for
  any $0 < \delta \le \delta(\gamma, \alpha) = 2\frac{\log( (1-\gamma )^{-2})-1}{\log(A)}$
   with $A = \frac{1-\alpha}{\alpha} > 1$. Further,
  $\delta(\gamma, 1/2) = \gamma(2-\gamma)(1-\gamma)^{-2}.$
\end{lemma}
\begin{proof}
  The estimate for $\delta(\gamma, 1/2)$ follows immediately from the
  above theorem assuming $\gamma < 1/2$. In the case when $\alpha < 1/2$,
  solving $\rho^2 \defeq (1-\gamma)^2 = (A^{1/q} - A^{-1/q}) (A^{1-1/q} - A^{1/q-1})^{-1}$ for $q$
  gives, for $\gamma < 1 - A^{-1/2}$,
  \[
  \delta = q - 2 = \frac{2\log(A)}{\log(\frac{1+\rho^2A}{1+\rho^2/A})} - 2 \geq 2\frac{\log( (1-\gamma )^{-2})-1}{\log(A)}.
  \]
\qedendproof


\full{
\paragraph{Efron-Stein Decompositions.} Our proofs make use of the
\textit{Efron-Stein decomposition} which has useful properties akin to
Fourier decompositions.
\begin{definition}
  Let $f: \Omega^{(1)} \times \dotsb \times \Omega^{(n)} \rightarrow
  \R$ and $\mu$ a measure on $\prod \Omega^{(t)}$. The
  \emph{Efron-Stein decomposition} of $f$ with respect to $\mu$ is
  defined as $\{f_S\}_{S \subseteq \irange{n}}$ where
\begin{gather*}
  f_S(\rvvx) = \sum_{T \subseteq S} (-1)^{\card{S \setminus T}}
  \condEx{f(\rvvx')}{\rvvx'_T = \rvvx_T}.
\end{gather*}
\end{definition}
\begin{lemma}[Efron and Stein~\cite{EfronS81}, and Mossel~\cite{Mossel10}]
\label{lemma:esproperties}
Assuming $\{\Omega^{(t)}\}_t$ are independent, the Efron-Stein
decomposition $\{f_S\}_S$ of $f: \prod \Omega^{(t)} \rightarrow \R$
satisfies:
\begin{itemize}
\item $f_S(\vx)$ depends only on $\vx_S$,
\item for any $S, T \subseteq \irange{m}$, and $\vx_T \in \prod_{t
    \in T} \Omega^{(t)}$ such that $S \setminus T \neq \emptyset$,
  $\condEx{f_S(\rvvx')}{\rvvx'_T = \vx_T} = 0.$
\end{itemize}
\end{lemma}

We use the standard notion of influence and noisy influence as in
previous work.

\begin{definition}
  Let $f: \Omega^n \to \R$ be a a function on a probabiltity space.
  The \emph{influence} of the $1 \le \ordinal{i} \le n$ coordinate is
  \[ \Inf[i]{f} = \Ex{\mathsf{Var}_{\Omega_i}(f)}. \]
  Additionally, given a noise parameter $\gamma$, the \emph{noisy influence} of the $\ordinal{i}$
  coordinate is 
  \[ \Inf[i][1-\gamma]{f} = \Ex[]{\mathsf{Var}_{\Omega_i} (\noiseop_{1-\gamma} f)}.  \]
\end{definition}

\def\inflinof{\Inf[i][1-\gamma]{f}}
The following bounds on the noisy influence are handy for the analysis.
\begin{lemma}\label{lemma:influence_variance_bound}%
  For every $\gamma > 0$, natural numbers $i$ and $n$ such that
  $1 \le i \le n$, and every $f: \Omega^n \to \R$, \[\inflinof \le
  \mathsf{Var}(f).\]
\end{lemma}  

\begin{lemma}\label{lemma:total_noisy_influence_bound}%
\label{prop:totInfBound}%
For every $\gamma > 0$, and every $f: \Omega^n \to \R$, \[\sum_i \inflinof
\le \frac{\mathsf{Var}(f)}{\gamma}.\]
\end{lemma}

We introduce the notion of \emph{\CoinfTerm} between functions which
is a notion implicitly prevalent in modern \labelcover reductions,
either for noised or for analytically similar degree-bounded functions:
\[
\Coinf{f, g} \defeq \sum_{(i,j) \in \pi} \Inf[i][1-\gamma]{f} \Inf[j][1-\gamma]{g}.
\]
We note that our definition differs somewhat from the 
cross-influence, denoted `$\mathrm{XInf}$', used by Samorodnitsky and Trevisan~\cite{SamorodnitskyT09}.

}


%% file: general_hardness.tex
\section{A General Hardness Result}
\label{sec:dictatorshipTest}

In this section, we prove a general inapproximability result for OCSPs which,
similar to results for classic CSPs, permit us to deduce hardness of
approximation based on the existence of certain simple distributions.
The proof is via a scheme of reductions from \labelcover to OCSPs.
For an $m$-width predicate $\payoff$, we instantiate this scheme with a
distribution $\baseDist$ over $Q_1^{t} \times Q_2^{m-t}$; for some
parameters $t$, $Q_1$, and $Q_2$; to obtain a reduction to
OCSP($\payoff)$ instances.
Straightforward applications of this result using
appropriate distributions yields \cref{thm:mbtw_result,thm:mnbtw_result,thm:mocsp_result}.

The reduction itself is composed of pieces known as dictatorship test
which is described in the next section.  \cref{sec:rednFromLC} uses
this test to construct the overall reduction and also contains the
properties of this reduction. Throughout this section, we assume
$\payoff$ is the $m$-width predicate of interest and that $\baseDist$
is the distribution of the appropriate signature.

\subsection{Dictatorship Test}\label{sec:generalDictatorshipTest}

The dictatorship test uses a distribution parametrized by $\gamma$,
and $\pi$ and is denoted by $\dictDist[\gamma]_\pi(\baseDist)$; its
definition follows.
 

\vspace{0.5\baselineskip} \hrule
\vspace{-0.4\baselineskip}
\begin{procedure}[Test Distribution]\label{proc:dict}\ 
  \vspace{-0.7\baselineskip}
  \paragraph{\textbf{Parameters:}}  
  \begin{itemize}
  \vspace{-0.9\baselineskip}
  \item distribution $\baseDist$ over
    $\overbrace{Q_1 \times \ldots \times Q_1}^{t} \times
    \overbrace{Q_2 \times \ldots \times Q_2}^{m-t}$;
  \item noise probability, $\gamma > 0$;
  \item projection map $\pi: \lcRightLabelSet \to \lcLeftLabelSet$;
  \end{itemize}
  \vspace{-\baselineskip}
  \paragraph{\textbf{Output:}} Distribution
  $\dictDist[\gamma]_{\pi}(\baseDist)$ over
  \[
	(\x[1], \dotsc, \x[t],
	\y[t+1], \dotsc, \y[m]) \in \left(Q_1^L \times \dotsm \times
    Q_1^L\right) \times \left(Q_2^R \times \dotsm \times
    Q_2^R\right).
  \]
  \begin{enumerate}
  \item pick a random $\card{\lcLeftLabelSet} \times t$ matrix $\X$
    over $Q_1$ by letting each row be a sample from $\baseDist_{\le
      t}$, independently.
  \item pick a random $\card{\lcRightLabelSet} \times (m-t)$ matrix
    $\Y \defeq (\y[t+1], \dotsc, \y[m])$ over $Q_2$ by letting the
    \ordinal{$i$} row be a sample from $\baseDist_{> t}$ conditioned on
    $\baseDist_{\le t} = \X_{\pi(i)} = (\super{\rvx}{1}_{\pi(i)}, 
    \dotsc, \super{\rvx}{t}_{\pi(i)})$.
  \item for each entry of $\X$ (resp.~$\Y$) independently, replace it
    with a sample from $Q_1$ (resp.~$Q_2$) with probability
    $\gamma$.
  \item output $(\X, \Y)$.
  \end{enumerate}
\end{procedure}
\hrule\vspace{\baselineskip}

\def\dicD{\dictDist[\gamma]_{\pi}(\baseDist)}
\def\dicDpie{\dictDist[\gamma]_{\pi_e}(\baseDist)}
\def\dicDP{\dictDist[\gamma]_{\pi}(\decoup{\baseDist})}
\def\dicDPpie{\dictDist[\gamma]_{\pi_e}(\decoup{\baseDist})}

Recall our convention from \cref{sec:ocsp prelims} of extending
$\payoff$ to a predicate $\payoff: \Z^m \to [0, 1]$. 
For a pair of functions $(f, g)$, let $(f,g) \circ (\X,\Y)$ denote
the tuple $(f(\x[1]), \ldots,f(\x[t])$, $g(\y[t+1]), \ldots, g(\y[m]))$. 
Then, the \emph{acceptance probability} on $\dicD$ for a pair of functions
$(f, g)$ where $f: Q_1^\lcLeftLabelSet \to \Z$ and $g: Q_2^\lcRightLabelSet \to \Z$
is:
\begin{equation}\label{eq:theProb}
  \accP(\dicD) \defeq \E_{(\X, \Y) \from \dicD}[\payoff((f,g) \circ (\X,\Y))].
\end{equation}
This definition is motivated by the overall reduction described in the
next section.  The distribution is designed so that functions $(f, g)$
that are dictated by a single coordinate have a high acceptance
probability, justifying the name of the test.

\begin{lemma}
  \label{lemma:dicttest completeness}
  Let $g: Q_2^\lcRightLabelSet \rightarrow \Z$ and $f: Q_1^\lcLeftLabelSet \rightarrow \Z$ be
  defined by $g(\vy) = y_i$ and $f(\rvx) = x_{\pi(i)}$ for some $i \in R$.  Then,
  $\accP(\dicD) \ge \E_{\x \sim \baseDist}[\payoff(\bfx)] - \gamma m$.
\end{lemma}
\begin{proof}
  The vector $(f(\x[1]), \ldots, f(\x[t]), g(\y[t+1]), \ldots,
  g(\y[m]))$ simply equals the \ordinal{$\pi(i)$} row of $\X$ followed by
  the \ordinal{$i$} row of $\Y$.  With probability $(1-\gamma)^m \ge 1-\gamma
  m$ this is a sample from $\baseDist$ and is hence accepted by
  $\payoff$ with probability at least $\E_{\rvvx \sim
    \baseDist}[\payoff(\rvvx)] - \gamma m$.
\qedendproof

We prove a partial converse of the above: unless $f$ and $g$ have
influential coordinates $i$ and $j$ such that $\pi(j) = i$, the
distribution $\baseDist$ can be replaced by a product of two
distributions with a negligible loss in the acceptance probability.
We define this product distribution below and postpone the analysis to
\cref{sec:test_soundness} in order to complete the description of the
reduction.

\begin{definition}
  Given the base distribution $\baseDist$, the \emph{decoupled}
  distribution $\decoup{\baseDist}$ is obtained by taking 
  independent samples $\rvvx$ from $\baseDist_{\le t}$ and $\rvvy$ from
  $\baseDist_{> t}$.
\end{definition}

\subsection{Reduction from Label Cover}\label{sec:rednFromLC}

\def\redgeneral{\super{R}{\payoff}_{\baseDist,\gamma}}
\def\redgeneralD{\super{R}{\payoff}_{\decoup{\baseDist},\gamma}}
\vspace{0.5\baselineskip}
\hrule
\vspace{0.2\baselineskip}
\begin{procedure}[Reduction $\redgeneral$]\label{proc:general reduction}\ 
  \vspace{-0.7\baselineskip}%
  \paragraph{\textbf{Parameters:}} distribution $\baseDist$ over $Q_1^t
  \times Q_2^{m-t}$ and noise parameter $\gamma > 0$.
  \vspace{-0.7\baselineskip}%
  \paragraph{\textbf{Input:}} a Label Cover instance $\lcDesc$.
  \vspace{-0.7\baselineskip}%
  \paragraph{\textbf{Output:}} a weighted $\ocsp(\payoff)$ instance $\instance
  = (\outputV, \outputE)$ where $\outputV = (\lcLeft \times Q_1^L)
  \cup (\lcRight \times Q_2^R)$.  The distribution of constraints in
  $\outputE$ is obtained by sampling a random edge $e = (u,v) \in E$
  with projection $\pi_e$ and then $(\X, \Y)$ from
  $\dictDist[\gamma]_{\pi_e}(\baseDist)$; the constraint is the
  predicate $\payoff$ applied on the tuple \[((u, \x[1]), \ldots, (u, \x[t]),
  (v, \y[t+1]), \ldots, (v, \y[m])).\]
\vspace{-0.9\baselineskip}
\hrule
\vspace{0.5\baselineskip}
\end{procedure}
\def\infl{\Inf[l][1-\gamma]}

An assignment to $\instance$ is seen as a collection of functions,
$\{f_u\}_{u \in U} \cup \{g_v\}_{v \in V}$, where $f_u: Q_1^L \to \Z$
and $g_v: Q_2^R \to \Z.$ The value of an assignment is:
\[
\E_{\stackrel{\rv{e} = (u,v) \in E;}{(\X, \Y) \in
    \dictDist[\gamma]_{\pi_{\rv{e}}}(\baseDist)}}
\payoff\left((f_u,g_v) \circ (\X,\Y)
\right) = \E_{\rv{e} = (u,v) \in E} [
\accP[f_u,g_v](\dictDist[\gamma]_{\pi_{\rv{e}}}(\baseDist)) ].
\]
\cref{lemma:dicttest completeness} now implies that if $\lc$ is
satisfiable, then the value of the instance output is also high.
\begin{lemma}\label{lemma:reduction completeness}%
  If $\lambda$ is a labeling of $\lc$ satisfying a fraction $c$ of
  its constraints, then the ordering assignment $f_u(\vx) =
  x_{\lambda(u)}$, $g_v(\vy) = y_{\lambda(v)}$ satisfies at least
  a fraction $c \cdot (\E_{\rvvx \sim
    \baseDist}[\payoff(\perm\sigma(\rvvx))] - \gamma m)$ of the
  constraints of $\redgeneral(\lc)$.  In particular, there is an
  ordering of $\redgeneral(\lc)$ attaining a value $\val(\lc) \cdot
  (\E_{\rvvx \sim \baseDist}[\payoff(\perm\sigma(\rvvx))] - \gamma m)$
  that is \emph{oblivious} to the distribution $\baseDist.$
\end{lemma}

On the other hand, we also extend the decoupling property of the
dictatorship test to the instance output if $\val(\lc)$ is small.
This is the technical core of the paper and is proved in
\cref{sec:proof of soundness}.

\begin{theorem}
  \label{theorem:reduction soundness}
  Suppose that $\baseDist$ over $Q_1^t \times Q_2^{m-t}$ satisfies the following properties:
  \begin{itemize}
  \item $\baseDist$ has uniform marginals.
  \item For every $i > t$, $\baseDist_i$ is independent of $\baseDist_{\le t}$.
  \end{itemize}
  Then, for every $\epsilon > 0$ and $\gamma > 0$ there exists
  $\epsilon_{LC} > 0$ such that if $\val(\mathcal{L}) \le
  \epsilon_{LC}$ then for every assignment $A = \{f_u\}_{u \in U} \cup
  \{g_v\}_{v \in V}$ to $\instance$ it holds that
  \[
  \val(A; \redgeneral(\mathcal{L})) \le \val(A;
  \redgeneralD(\mathcal{L})) + \epsilon.
  \]
  In particular, $\val(\redgeneral(\mathcal{L})) \le
  \val(\redgeneralD(\mathcal{L})) + \epsilon$.
\end{theorem}


%% file: applications.tex
\def\ocspEpsilon{\ensuremath{\epsilon_{\text{s}}}}
\def\lcEpsilon{\ensuremath{\epsilon_{\text{LC}}}}

\section{Applications of the General Result}\label{sec:inapproxRedn}

\full{
In this section, we prove the inapproximability of
\MaxBTW, \MaxNBTW, and \MaxSO, using the
general hardness result of \cref{sec:dictatorshipTest}. We also
prove the hardness of \MAS using a
gadget reduction from \MaxNBTW.
}
\lncs{
In this section, we prove the inapproximability of
\MaxBTW and \MaxNBTW using the
general hardness result of \cref{sec:dictatorshipTest}. We also
prove the hardness of \MAS using a
gadget reduction from \MaxNBTW.  For space concerns, the
inapproximability of \MaxSO is deferred to the full version.
}


\def\dbtwSpaceOne{ \{ -1,q\} }
\def\dbtwSpace{ \dbtwSpaceOne \times \irange{q} \times \irange{q} }
\def\dbtw{\baseDist}
\def\ddbtw{\decoup{\baseDist}}
\def\redbtw{R^\BTW_{\dbtw,\gamma}}
\def\dredbtw{R^\BTW_{\ddbtw,\gamma}}

\def\redbtwIns{\redbtw(\lc)}
\def\dredbtwIns{\dredbtw(\lc)}
\def\mbtwEpsilon{\ocspEpsilon}

\subsection{Hardness of \textproblem{Maximum Betweenness}}\label{sec:MBTW}
For an integer $q$, define the distribution
$\dbtw$ over $\dbtwSpace$ by picking $\rvx_1 \sim \dbtwSpaceOne$, $\rvy_2
\sim \irange{q}$
, and setting $\rvy_3 = \rvy_2 + 1 \bmod q\text{ if $\rvx_1 = q$ and }
\rvy_2 - 1 \text{ otherwise.}$
This distribution has the following properties which can
be readily verified.
\begin{proposition}\label{prop:mbtw_dist_props}
Let $(\rvx_1, \rvy_2, \rvy_3) \sim \dbtw$. Then the following holds:
\begin{enumerate}
\item $\dbtw$ has uniform marginals.
\item The marginals $\rvy_2$ and $\rvy_3$ are independent of $\rvx$.
\item $(\rvy_2, \rvy_3)$ has the same distribution as $(\rvy_3, \rvy_2)$.
\item \label{prop:btw distr complete} $\E_{\rvx_1, \rvy_2, \rvy_3 \sim \dbtw}[\BTW(\rvx_1, \rvy_2, \rvy_3)] \ge 1 - 1/q$.
\end{enumerate}
\end{proposition}

Let $\ddbtw$ be the decoupled distribution of $\dbtw$ which draws the
first coordinate independently of the remaining and $\gamma > 0$ a
noise parameter.  Given a \labelcover instance $\mathcal{L}$ and
consider applying Reduction~\ref{proc:general reduction} to $\mathcal{L}$
with test distributions $\dbtw$ and $\ddbtw$, obtaining \MaxBTW instances
$\instance = \redbtwIns$ and $\instance^{\perp} = \dredbtwIns$.

\begin{lemma}[Completeness]
  \label{lemma:mbtw completeness}
  If $\val(\mathcal{L}) = 1$ then $\val(\instance) \ge 1 - 1/q - 3\gamma$.
\end{lemma}

\begin{proof}
  This is an immediate corollary of
  \cref{lemma:reduction completeness,prop:mbtw_dist_props}.
\qedendproof

\begin{lemma}[Soundness]
  \label{lemma:mbtw soundness}
  For every $\epsilon > 0$, $\gamma > 0$, $q$, there is an $\epsilon_{LC} >
  0$ such that if $\val(\mathcal{L}) \le \epsilon_{LC}$ then $\val(\instance)
  \le 1/2 + \epsilon$.
\end{lemma}

\begin{proof}
  We note that
  \cref{prop:mbtw_dist_props} asserts that $\dbtw$ satisfies the
  conditions of \cref{theorem:reduction soundness} and it
  suffices to show $\val(\instance^{\perp}) \le 1/2$.  Let $\{f_u:
  \{0,1\}^\lcLeftLabelSet \rightarrow \Z \}_{u \in \lcLeft}, \{g_v:
  \irange{q}^\lcRightLabelSet \rightarrow \Z\}_{v \in \lcRight}$ be an
  arbitrary assignment to $\instance^{\perp}$. Fix an \labelcover edge
  $\{u,v\}$ with projection $\pi$ and consider the mean value of
  constraints produced for this edge by the construction:
  \begin{gather}
    \label{eq:3490857t89wh}
    \Ex[\super{\x}{1}, \super{\rvvy}{2}, \super{\rvvy}{3} \leftarrow
    \ntestDist{\ddbtw}]{\BTW(f_u(\x[1]), g_v(\y[2]), g_v(\y[3]))}.
  \end{gather}
  
  As noted in \cref{prop:mbtw_dist_props}, $(\y[2], \y[3])$ has the
  same distribution as $(\y[3],\y[2])$ when drawn from
  $\dbtw$. Consequently, when drawing arguments from the
  decoupled test distribution, the probability of a specific outcome
  $(\super{\vx}{1}, \super{\vy}{2}, \super{\vx}{3})$ equals the
  probability of $(\super{\vx}{1}, \super{\vy}{3}, \super{\vx}{2})$. For
  strict orderings, at most one of the two can satisfy the predicate
  $\BTW$.  Thus, the expression in $(\ref{eq:3490857t89wh})$, and in effect
  $\val(\instance^{\perp})$, is bounded by $1/2$.
\qedendproof

\cref{thm:mbtw_result} is now an immediate corollary of
\cref{lemma:mbtw completeness,lemma:mbtw soundness},
taking $q = \lceil 2/\epsilon \rceil$ and $\gamma = \epsilon/6$.


\def\dnbtw{\baseDist_{3}}
\def\dnbtwn{\baseDist_{4}}
\def\dnbtwp{\dnbtwn}
\def\ddnbtw{\decoup{\baseDist}_{3}}
\def\ddnbtwn{\decoup{\baseDist}_{4}}
\def\ddnbtwp{\ddnbtwn}
\def\rednbtw{R^{\NBTW}_{\dnbtw{}, \gamma}}
\def\rednbtwn{R^{\NBTW}_{\dnbtwn{}, \gamma}}
\def\rednbtwp{R^{\NBTW'}_{\dnbtwp{}, \gamma}}
\def\rednbtwc{R^{\NBTW}_{\text{comp}, \gamma}}
\def\drednbtw{R^{\NBTW}_{\ddnbtw{}, \gamma}}
\def\drednbtwn{R^{\NBTW}_{\ddnbtwn{}, \gamma}}
\def\rednbtwIns{\rednbtw(\lc)}
\def\rednbtwInsn{\rednbtwn(\lc)}
\def\rednbtwInsp{\rednbtwp(\lc)}
\def\rednbtwInsc{\rednbtwc(\lc)}
\def\drednbtwIns{\drednbtw(\lc)}
\def\drednbtwInsn{\drednbtwn(\lc)}

\def\mnbtwEpsilon{\ocspEpsilon}

\def\dnbtw{\baseDist}
\def\ddnbtw{\decoup{\baseDist}}

\subsection{Hardness of \textproblem{Maximum Non-Betweenness}}\label{sec:MNBTW}

For an implicit parameter $q$, define a distribution $\dnbtw$ over
$[q]^3$ by picking $\rvx_1, \rvx_2 \sim [q]$ and setting $\rvx_3 =
\rvx_1 + \rvx_2 \bmod q$.

\begin{proposition}\label{prop:mnbtw_dist_props}%
  The distribution $\dnbtw$ satisfies the following:
  \begin{enumerate}
  \item $\dnbtw$ is pairwise independent with uniform marginals,
  \item and $\Ex[\bfx_1, \bfx_2, \bfx_3 \sim \dnbtw]{\NBTW(\bfx_1, \bfx_2, \bfx_3)} \ge 1 - 3/q$.
  \end{enumerate}
\end{proposition}

A straightforward application of the general inapproximability with $t
= 1$ shows that $\rvx_1$ is decoupled from $\rvx_2$ and $\rvx_3$
unless $\val(\lc)$ is large.  Further, pairwise independence implies
that the decoupled distribution is simply the uniform distribution
over $[q]^3.$ However, this does not suffice to prove approximation
resistance and in fact the value could be greater than $2/3$.  To see this,
note that if $\set{f_u}_{u \in U}, \set{g_v}_{v \in V}$ is an ordering
of the instance from the reduction, then the first coordinate
of every constraint is a variable of the form $f_u(\cdot)$ while the
rest are $g_v(\cdot)$. Thus, ordering the $f_u(\cdot)$ variables in
the middle and randomly ordering $g_v(\cdot)$ on both sides satisfies
a fraction $3/4$ of the constraints.

To remedy this and prove approximation
resistance, we permute $\dnbtw$ by swapping the last coordinate with
each of the remaining coordinates and overlay the instances obtained by the
reduction obtained from these respective distributions.
More specifically, for $1 \le j \le 3$, define $\dnbtw_j$ as the distribution over
$[q]^3$ obtained by first sampling from $\dnbtw$ and then swapping the
\ordinal{$j$} and third coordinate, \ie, the \ordinal{$j$} coordinate is
the sum of the other two which are picked independently at random.
Similarly, define $\NBTW_j$ as the ordering predicate which is true if
the \ordinal{$j$} argument does not lie between the other two, \eg, $\NBTW_3 =
\NBTW$.

As in the previous section, take a \labelcover instance $\mathcal{L}$
and consider applying Reduction~\ref{proc:general reduction} to
$\mathcal{L}$ with the distributions $\dnbtw_j$, and write
$\instance_j = R^{\NBTW_j}_{\dnbtw_j, \gamma}(\mathcal{L})$.  Similarly write
$\instance_j^{\perp} = R^{\NBTW_j}_{\ddnbtw_j, \gamma}(\mathcal{L})$ for the
corresponding decoupled instances.

As the distributions $\dnbtw_j$ are over the same domain $[q]^3$, the
instances $\instance_1, \instance_2, \instance_3$ are over the same
variables.  We define a new instance $\instance$ over the same
variables as the ``sum'' $\frac{1}{3} \sum_{j \in [3]} \instance_j$,
defined by taking all constraints in $\instance_1, \instance_2,
\instance_3$ with multiplicities and normalizing their weights
by $1/3.$

\begin{lemma}[Completeness]
  \label{lemma:mnbtw completeness}
  If $\val(\mathcal{L}) = 1$ then $\val(\instance) \ge 1 - 3/q - 3\gamma$.
\end{lemma}

\begin{proof}
  This is an immediate corollary of \cref{lemma:reduction
    completeness} and \cref{prop:mnbtw_dist_props}.
\qedendproof

\begin{lemma}[Soundness]
  \label{lemma:mnbtw soundness}
  For every $\epsilon > 0$, $\gamma > 0$, $q$, there is an $\epsilon_{LC} >
  0$ such that if $\val(\mathcal{L}) \le \epsilon_{LC}$ then $\val(\instance)
  \le 2/3 + \epsilon$.
\end{lemma}

\begin{proof}
  Again our goal is to use \cref{theorem:reduction soundness} and
  we start by bounding $\val(\instance^\perp)$.  To do this, note that
  the decoupled distributions $\dnbtw_j$ are in fact the uniform
  distribution over $[q]^3$ and in particular do not depend on $j$.
  This means that the distributions of variables which $\NBTW_j$ is
  applied to in $\instance_j^\perp$ is independent of $j$, \eg, if
  $\instance_1^\perp$ contains the constraint $\NBTW_1(z_1, z_2, z_3)$
  with weight $w$ then $\instance_2^\perp$ contains the constraint
  $\NBTW_2(z_1, z_2, z_3)$ with the same weight.  In other words,
  $\instance^{\perp}$ can be thought of as having constraints of the
  form $\Ex[j]{\NBTW_j(z_1, z_2, z_3)}$.  It is readily verified that
  $\Ex[j]{\NBTW_j(a,b,c)]} \le 2/3$ for every $a, b, c$.

  Getting back to the main task -- bounding $\val(\instance)$ --
  fix an arbitrary assignment $A = \{f_u: \irange{q}^L \}_{u \in
    \lcLeft } \cup \{g_v: \irange{q}^R \}_{v \in \lcRight}$ of
  $\instance$.  By \cref{theorem:reduction soundness}, $\val(A;
  \instance_j) \le \val(A; \instance_j^\perp) + \epsilon$ for $j \in
  [3]$.  It follows that $\val(A; \instance) \le \val(A;
  \instance^\perp) + \epsilon$ and therefore, since $A$ was arbitrary, it holds that $\val(\instance)
  \le \val(\instance^\perp) + \epsilon \le 2/3 + \epsilon$, as desired.
\qedendproof

\subsection{Hardness of \textproblem{Maximum Acyclic Subgraph}}
\label{sect:mas_result}

\full{
The inapproximability of \MAS is from a simple gadget reduction from
\MaxNBTW.  We claim the following properties of the
directed graph shown in~\cref{fig:mas_section_mas_gadget}, defined
formally as follows.

\begin{definition}
\label{def:mas_gadget}
Define the \MAS gadget $H$ as the directed graph $H
\defeq (V, A)$ where $V = \{x, y, z,
a, b\}$ and $A$ consists of the walk
\[
b \rightarrow x \rightarrow a \rightarrow z
\rightarrow b \rightarrow y \rightarrow a.
\] 
\end{definition}
}

\lncs{
The inapproximability of \MAS is from a simple gadget reduction from
the inapproximability gap of \MaxNBTW.  We claim the following properties of the
directed graph shown in~\cref{fig:mas_section_mas_gadget}.  The proof and the routine application of the lemma to derive \cref{thm:mas_result} are given in the full version.
}

\begin{lemma} 
  \label{prop:mas gadget value}
  Consider an ordering $\ordr$ of $x$, $y$, $z$.  Then, \begin{enumerate}
  \item if $\NBTW(\ordr(x), \ordr(y), \ordr(z)) = 1$, then
    $\max_{\ordr'} \val(\ordr'; H) = 5/6$ where the $\max$ is over all
    extensions $\ordr': V \to \Z$ of $\ordr$ to $V.$
  \item if $\NBTW(\ordr(x), \ordr(y), \ordr(z)) = 0$, then
    $\max_{\ordr'} \val(\ordr'; H) = 4/6$ where the $\max$ is over all
    extensions of $\ordr$ to $V.$
  \end{enumerate}
\end{lemma}

\full{
\begin{proof}
  To find the value of the gadget $H$, we individually consider the
  optimal placement of $a$ and $b$ relative $x,y,z$.  There are three
  edges in which the respective variables appear: $a$ appears in $(x,
  a)$, $(y, a)$ and $(a, z)$; while $b$ appears in $(z, b)$, $(b, x)$,
  and $(b, y)$.

  From this, we gather that two out of the three respective
  constraints can always be satisfied by placing $a$ after $x,y,z$ and
  similarly placing $b$ before $x,y,z$. We also see that all three
  constraints involving $a$ can be satisfied if and only if $z$ comes
  after both $x$ and $y$. Similarly, satisfying all three constraints
  involving $b$ is possible if and only if when $z$ comes before both
  $x$ and $y$. From this, one concludes that if $\NBTW(x, y, z) = 1$,
  i.e., if $z$ comes first or last, then we can satisfy five out of the
  six constraints, whereas if $z$ is the middle element of $\ordr$, we
  can satisfy only four out of the six constraints.
\qedendproof

The proof of \cref{thm:mas_result} is now a routine application
of the \MAS gadget.

\begin{proof}[Proof of \cref{thm:mas_result}]
  Given an instance $\instance$ of \MaxNBTW, construct a directed graph
  $G$ by replacing each constraint $\NBTW(x,y,z)$ of $\instance$ with
  a \MAS gadget $H$, identifying $x,y,z$ with the vertices $x,y,z$ of
  $H$ and using two new vertices $a,b$ for each constraint of
  $\instance$.

  By \cref{prop:mas gadget value}, it follows that $\val(G) =
  \frac{5}{6}\val(\instance) + \frac{4}{6}(1-\val(\instance))$.  By
  \cref{thm:mnbtw_result}, it is \NP-hard to distinguish between
  $\val(\instance) \ge 1-\epsilon$ and $\val(\instance) \le
  2/3+\epsilon$ for every $\epsilon > 0$, implying that it is
  $\NP$-hard to distinguish $\val(G) \ge 5/6 - \epsilon/6$
  from $\val(G) \le 7/9 + \epsilon/6$, providing a hardness gap of
  $\frac{7/9}{5/6} + \epsilon' = 14/15 + \epsilon'$.
\qedendproof
}


\def\dmocsp{\baseDist}
\def\ddmocsp{\decoup{\baseDist}}
\def\redmocsp{R^{\SOPred[2t]}_{\dmocsp{}, \gamma}}
\def\dredmocsp{R^{\SOPred[2t]}_{\ddmocsp{}, \gamma}}
\def\redmocspIns{\redmocsp(\lc)}
\def\dredmocspIns{\dredmocsp(\lc)}
\def\mocspEpsilon{\ocspEpsilon}

\subsection{Hardness of \textproblem{Maximum $2t$-Same Order}}
\newcommand{\ind}{\mathrm{\mathbf{1}}}
We establish the hardness of \MaxSO,
\cref{thm:mocsp_result}, via the approximation resistance of the
relatively sparse predicate \SOPred[2t]. The proof is similar to the that of
\MaxBTW~(see \cref{sec:MBTW}).  Let $q_1 < q_2$ be integer parameters and
define the base distribution $\baseDist$ over $\irange{q_1}^t \times
\irange{q_2}^t$ as follows: draw $\bfx_1, \ldots, \bfx_t$ uniformly at
random from $\irange{q_1}$, draw $\bfz$ uniformly at random from
$[q_2]$, and for $1 \le j \le t$ set $\bfy_j = \bfx_j + \bfz \bmod
q_2$.  The distribution of $(\bfx_1, \ldots, \bfx_t, \bfy_1, \ldots,
\bfy_t)$ defines $\baseDist$.  For a permutation $\perm\sigma \in S_t$, let
$\ind_\sigma(\cdot)$ be the ordering predicate which is $1$ on $\perm\sigma$ and
$0$ on all other inputs.

\begin{proposition}\label{prop:mocsp_dist_props}
$\baseDist$ satisfies the following properties.
\begin{enumerate}
\item $\baseDist$ has uniform marginals.
\item For every $i > t$, $\baseDist_i$ is independent of $\baseDist_{\le t}$.
\item\label{prop:mocsp dist soundness} For every $\perm\sigma \in S_t$,
$\Ex{\ind_\sigma(\rvx_1, \ldots, \rvx_t)} = \Ex{\ind_\sigma(\rvy_{t+1}, \ldots, \rvy_m)} = 1/t!$
\item\label{prop:mocsp dist completeness} $\Ex[\rvx_1, \ldots, \rvx_t, \rvy_{t+1}, \ldots, \rvy_m \sim \baseDist]{\SOPred[2t](\rvx_1, \ldots, \rvx_t, \rvy_{t+1}, \ldots, \rvy_m)} \ge 1 - \frac{t^2}{2q_1} - \frac{q_1}{q_2}$.
\end{enumerate}
\end{proposition}

\Ctodo{We are not being consistent in sub vs superindexing}
\begin{proof}
  The first three properties are immediate from the construction and recalling the extension of predicates to non-unique values.
  For the last property, note that $\SOPred[2t](\rvx_1, \ldots, \rvx_t, \rvy_{t+1}, \ldots, \rvy_m) = 1$ if $\rvx_1, \ldots, \rvx_t$ are distinct
  and $\rvz < q_2 - q_1$. The former event occurs with probability at least $1 - \frac{t^2}{2q_1}$ and the latter independently with  probability at least $1 - q_1/q_2$; a union bound implies the claim. 
\qedendproof

As in the proof of \cref{thm:mbtw_result}, take a \labelcover instance
$\mathcal{L}$ and let $\instance = \redmocsp(\mathcal{L})$ and
$\instance^{\perp} = \dredmocsp(\mathcal{L})$ be the instances
produced by Reduction~\ref{proc:general reduction} using the base distribution $\baseDist$ and the decoupled version $\baseDist^{\perp}$ and some noise parameter $\gamma > 0$.

The following lemma is an immediate corollary of
\cref{lemma:reduction completeness} and
\cref{prop:mocsp_dist_props}, Item~\ref{prop:mocsp dist completeness}.

\begin{lemma}[Completeness]
  \label{lemma:mocsp completeness}
  If $\val(\mathcal{L}) = 1$ then $\val(\instance) \ge 1 - \frac{t^2}{2q_1} - \frac{q_1}{q_2} - 3\gamma$.
\end{lemma}

For the soundness, we have the following.

\begin{lemma}[Soundness]
\label{lemma:mocsp soundness}
For every $\epsilon > 0, \gamma > 0, $ and $1 \le q_1 \le q_2$, there is an
$\epsilon_{LC} > 0$ such that if $\val(\mathcal{L}) \le \epsilon_{LC}$
then $\val(I) \le \frac{1}{t!} + \epsilon$.
\end{lemma}

\begin{proof}
  As in the proof of \cref{lemma:mbtw soundness}, it suffices to
  prove $\val(\instance^{\perp}) \le 1/t!$. %
  Let $\{f_u: \irange{q_1}^L \rightarrow \Z\}_{u \in \lcLeft}$,
  $\{g_v: \irange{q_2}^R \rightarrow \Z\}_{v \in \lcRight}$ be an
  arbitrary assignment to $\instance^{\perp}$.  Fix an arbitrary edge
  $\{u, v\}$ of $\mathcal{L}$ with projection $\pi$.  The value of
  constraints corresponding to $\{u,v\}$ satisfied by the assignment
  is
\begin{align*}
  \lefteqn{\Ex[(\X, \Y) \in \ntestDist{\ddmocsp}]{\SOPred[2t](f_u(\x[1]), \ldots, f_u(\x[t]), g_v(\y[t+1]), \ldots, g_v(\y[m]))}} \hphantom{12345}\\
  &= \sum_{\sigma \in S_t} \Ex[(\X, \Y) \in \ntestDist{\ddmocsp}]{ \iverson[\sigma]{f_u(\x[1]), \ldots, f_u(\x[t])} \iverson[\sigma]{g_v(\y[t+1]), \ldots, g_v(\y[m])}} \\
  &= \sum_{\sigma \in S_t} \Ex[(\X, \Y) \in \ntestDist{\ddmocsp}]{ \iverson[\sigma]{f_u(\x[1]), \ldots, f_u(\x[t])}}  \Ex[(\X, \Y) \in \ntestDist{\ddmocsp}]{ \iverson[\sigma]{g_v(\y[t+1]), \ldots, g_v(\y[m])}} \\
  & = 1/t!,
\end{align*}
where the penultimate step uses the independence of $\X$ and $\Y$ in
the decoupled distribution, and the final step
Item~\ref{prop:mocsp dist soundness} of \cref{prop:mocsp_dist_props}.
\qedendproof

\cref{thm:mocsp_result} is an immediate corollary of
\cref{lemma:mocsp completeness,lemma:mocsp soundness}, taking $q_1 =
\lceil 2t^2/\epsilon \rceil$, $q_2 = \lceil 3q_1/\epsilon \rceil$ and
$\gamma = \epsilon/9$.


%% file: general_soundness_proof.tex
\section{Analysis of the Reduction}
\label{sec:proof of soundness}

In this section we prove \cref{theorem:reduction soundness} which
bounds the value of the instance generated by the reduction in terms of
the decoupled distribution.  Throughout, we fix an LC instance $\lc$,
a predicate $\payoff$, an OCSP instance $\instance$ obtained by
the procedure $\redgeneral$ for a distribution $\baseDist$ and
noise-parameter $\gamma$, and finally an assignment
$A = \set{f_u}_{u \in  \lcLeft} \cup \set{g_v}_{v \in \lcRight}$.

The proof involves three major steps.
First, we show that the assignment functions, which are $\Z$-valued, can be
approximated by functions on finite domains via bucketing~(see
\cref{sec:bucketing}).  This approximation makes the analyzed
instance value suspectible to tools
developed in the context of finite-domain
CSPs~\cite{Wenner12,Chan13} which are used in
\cref{sec:test_soundness} to prove the decoupling property of the
dictatorship test.  Finally, this decoupling is extended to the
reduction hence bounding the value of
\ifdefined\fullVersion
$\instance$ (see \cref{sec:reduction soundness}).
\else
$\instance$.
\fi

\def\buckF{B^{(f)}}
\def\buckG{B^{(g)}}
\def\buck{\bucketPred^{(f,g)}}
\def\extF{R^{(f)}}
\def\extG{R^{(g)}}
\def\buckParameter{\Gamma}

\subsection{Bucketing}\label{sec:bucketing}

\def\buckF{B^{(f_u)}} \def\buckG{B^{(g_v)}}%
For an integer $\buckParameter$, we approximate the function $f_u:
Q_1^L \to \Z$ by partitioning the domain into $\buckParameter$ pieces.
Set $q_1 = \card{Q_1}$ and partition the set $Q_1^L$ into sets
$\buckF_1,\ldots,\buckF_{\buckParameter}$ of size
$q_1^L/{\buckParameter}$ such that if $\vx \in \buckF_i$ and $\vy \in
\buckF_j$ for some $i < j$ then $f(\vx) < f(\vy)$.  Note that this is
possible as long as the parameter $\buckParameter$ divides $q_1^L$
which will be the case.  Let $\bucketInd[]_u: Q_1^L \to
[{\buckParameter}]$ specify the mapping of points to the bucket
containing it, and $\bucketInd[a]_u: Q_1^L \rightarrow \{0,1\}$ the
indicator of points assigned to $\buckF_a$.  Partition $g_v: Q_2^R \to
\Z$ similarly into buckets $\{\buckG_{a}\}$ obtaining
$\bucketIndR[]_v: Q_2^R \to [{\buckParameter}]$ and $\bucketIndR[a]_v:
Q_2^R \to \bits.$ \def\buckF{B^{(f)}} \def\buckG{B^{(g)}}%

Now we show that the acceptance probability of the dictatorship test
-- see (\ref{eq:theProb}) in \cref{sec:dictatorshipTest} -- applied to
an edge $e = (u, v)$ of the LC instance $\lc$ can be approximated by a
bucketed version.  Fix an edge $e = (u, v)$ and put $f = f_u$, $g =
g_v$.  As before, we denote a query tuple,
$(\x[1],\ldots,\x[t],\y[t+1],\ldots,\y[m])$ concisely as $(\X, \Y)$
and the tuple of assignments by a pair of functions $(f, g)$,
$(f(\x[1]), \ldots,f(\x[t])$, $g(\y[t+1]),\ldots,g(\y[m]))$ as $(f,g)
\circ (\X,\Y).$ Define the bucketed payoff function with respect to
$f$ and $g$, $\buck: [{\buckParameter}]^m \to \bndI$ as:
\begin{align}
  \buck(a_1,\ldots,a_m) &= \mathop{\E_{\x[i] \from \buckF_{a_i}; i\le
      t}}_%
  {\y[j] \from \buckG_{a_j}; t<j} \left[\payoff((f,g) \circ (\X,\Y))
  \right]\label{eq:bucketPayoff}\\
  \intertext{and the \emph{bucketed} acceptance probability, }
  \bVal(\dicD) &= \Ex[(\X,\Y) \from \dicD]{ \buck(
    (F,G) \circ (\X, \Y))
  }.\label{eq:bucketVal}
\end{align}
In other words, bucketing corresponds to generating a tuple $\vec{a} = (f, g) \circ (\X, \Y)$
and replacing each coordinate $a_i$ with a random value from the bucket $a_i$ fell in. 
We show that above is close to the true acceptance
probability $\accP(\dicD)$.

\begin{theorem}\label{prop:bucketing_loss_bound}%
  For every predicate $\payoff$, every distribution
  $\baseDist$ with uniform marginals, every pair of orderings $f:
  Q_1^L \to \Z$ and $g: Q_2^R \to \Z$, every $\gamma > 0$, projection $\pi: R \to L$, and every ${\buckParameter}$,
  \[ \left|\accP(\dicD) -%
    \bVal(\dicD) \right| \le m^2 {\buckParameter}^{-\delta}, \]
    for some $\delta = \delta(\gamma, Q) > 0$ with $Q = \max \{\card{Q_1}, \card{Q_2}\}.$
\end{theorem}

To prove this, we show that $f$ and $g$ have few overlapping pairs of
buckets and that the probability of hitting any particular pair is
small. Let $\extF_a$ be the smallest interval in $\Z$ containing
$\buckF_a$; and similarly $\extG_a.$

\begin{lemma}[Few Buckets Overlap]\label{prop:bucket_overlap_bound}%
  For every integer ${\buckParameter}$ there are at most
  $2{\buckParameter}$ choices of pairs $(a, b) \in [{\buckParameter}]
  \times [{\buckParameter}]$ such that $\extF_{a} \cap \extG_{b} \ne
  \emptyset.$
\end{lemma}
\begin{proof}
  Construct the bipartite intersection graph $G_I = (U_I, V_I, E_I)$
  where the vertex sets are disjoint copies of $[{\buckParameter}]$,
  and there is an edge between $a \in U_I$ and $b
  \in V_I$ iff $\extF_a \cap \extG_{b} \ne \emptyset$. By
  construction of the buckets, the graph does not contain any pair of
  distinct edges $(u,v), (u',v')$ such that $u < v$ and $u' >
  v'$. Consequently, a vertex can have at most two neighbors with
  degree greater than one. Let $A$ be the set of degree-one
  vertices. It follows that the maximum degree of the subgraph
  induced by $\comp{A}$ is at most two and
  contains at most $\card{(U_I \cup V_I) \setminus A}$ edges. 
  On the other hand, the number of edges incident to $A$
  is at most $\card{A}$ implying a total of at most
  $\card{U_I \cup V_I} = 2\Gamma$ intersections.
\qedendproof

Next, we prove a bound on the probability that a fixed pair of the $m$
queries fall in a fixed pair of buckets.  For a distribution $\cD$ over
$Q_1^{\lcLeftLabelSet} \times Q_2^{\lcRightLabelSet}$, define
$\cD^{(\gamma)}$ as the distribution that samples from $\cD$ and for each
of the $\card{\lcLeftLabelSet} + \card{\lcRightLabelSet}$ coordinates
independently with probability $\gamma$ replaces it with a new sample
from $\cD$. The distribution $\cD^{(\gamma)}$ is representative of the projection of
$\dicD$ to two specific coordinates and we show that noise prevents
the buckets from intersecting with good probability.

\begin{lemma}\label{prop:hyperContractiveBound}%
  Let $\cD$ be a distribution over $Q_1^{\lcLeftLabelSet} \times Q_2^{\lcRightLabelSet}$
  whose marginals are uniform in $Q_1^{\lcLeftLabelSet}$ and $Q_2^{\lcRightLabelSet}$ and
  $\cD^{(\gamma)}$ be as defined above.  For every integer ${\buckParameter}$ and every
  pair of functions $F: Q_1^{\lcLeftLabelSet} \to \bits$ and $G: Q_2^{\lcRightLabelSet} \to
  \bits$ such that $\Ex{F(\x)} = \Ex{G(\y)} = 1/{\buckParameter}$,
  \[ \Ex[(\x,\y) \in \cD^{(\gamma)}]{ F(\x) G(\y) } \le {\buckParameter}^{-(1 + \delta)}\]
  for some $\delta = \delta(\gamma, Q) > 0$ where $Q = \min\{\card{Q_1}, \card{Q_2}\}$.
\end{lemma}
\def\nOp{T_{1-\gamma}}
\begin{proof} Without loss of generality, let $\card{Q_1} = \min\{\card{Q_1}, \card{Q_2}\}$.
Set $q = 2 + \delta' > 2$ as in \cref{prop:bestHCBound}, $1/q' = 1 - 1/q$,
  and define $H(\x ) \defeq \Ex[\y | \x ]{\nOp G(\y )}$. Then,
  \begin{align*}
    \Ex[(\x,\y) \in \cD^{(\gamma)}]{ F(\x) G(\y) }
    &= \Ex[\x]{ \nOp F(\x) H(\x) }
    \le \norm{ \nOp F }_q \norm { H }_{q'}  
    \\&\le \norm{F}_2 \norm{H}_{q'}
    = \norm{F}_2 \norm{\nOp G}_{q'} 
    \\&\le \norm{ \nOp F }_q \norm { G }_{q'}
    = {\buckParameter}^{-(1/2 + 1/q')}
    = {\buckParameter}^{-(1 + \nicefrac{\delta'}{2(2+\delta')})},
  \end{align*}
  using \cref{prop:bestHCBound}, convexity of norms, and the contractivity of $\nOp$.
\qedendproof

Note that the above lemma applies to queries to the same function as
well, setting $F = G$, etc.  To complete the proof of
\cref{prop:bucketing_loss_bound}, we apply the above lemma to every
distinct pair of the $m$ queries made in $\dicD$ and each of the (at
most) $2 \buckParameter$ pairs of overlapping bucketes, bounding the
difference between the true acceptance probability and the bucketed
version.

\begin{proof}[Proof of \cref{prop:bucketing_loss_bound}]
  Note that the bucketed payoff $\buck( 
  (F,G) \circ (\X,\Y))$ is equal to the true payoff $\payoff(
  (f,g) \circ (\X,\Y))$ except possibly when two pairs of outputs fall
  in an overlapping pair of buckets. 
  Fix a pair of inputs, say $\super{x}{i}$ and $\super{y}{j}$; the argument is the
  same if we choose two inputs from $\X$ or from $\Y$.  Let $a =
  F(\super{x}{i})$ and $b = G(\super{y}{j})$.  By \cref{prop:bucket_overlap_bound}
  there are at most $2{\buckParameter}$ possible values $(a,b)$ such that the buckets
  indexed by $a$ and $b$ are overlapping. From \cref{prop:hyperContractiveBound},
  the probability that $F(\x[i]) = a$ and $G(\y[j]) = b$ is at most
  ${\buckParameter}^{-1-\delta}$.  By a union bound, the two outputs $F(\x[i])$,
  $G(\y[j])$ consequently fall in overlapping buckets with probability at most
  $2{\buckParameter}^{-\delta}$.  As there are at most ${m \choose 2} \le m^2/2$ pairs
  of outputs, the proof is complete.
\qedendproof

\subsection{Soundness of the Dictatorship Test}\label{sec:test_soundness}

We now reap the benefits of bucketing and prove the decoupling
property of the dictatorship test alluded to in
\cref{sec:dictatorshipTest}.  

\full{
\begin{theorem}
  \label{lemma:test bucketed soundness}
  For every predicate $\payoff$ and distribution $\baseDist$
  satisfying the conditions of \cref{theorem:reduction soundness},
  and any noise rate $\gamma > 0$, projection $\pi: \lcRightLabelSet \to \lcLeftLabelSet$, and
  bucketing parameter ${\buckParameter}$, the following holds.  For any functions $f:
  Q_1^\lcLeftLabelSet \to \Z$, $g: Q_2^\lcRightLabelSet \to \Z$ with bucketing functions $F:
  Q_1^\lcLeftLabelSet \to [{\buckParameter}]$, $G: Q_2^\lcRightLabelSet \to [{\buckParameter}]$,
 \begin{align*}
  \abs{ \bVal(\dicD) -  \bVal(\dicDP) } \le \gamma^{-1/2}
 m^{1/2} 4^m {\buckParameter}^m \sum_{a, b \in \irange{{\buckParameter}}} \Coinf{\super{F}{a},
   \super{G}{b}}^{1/2}.
 \end{align*}
\end{theorem}

Recall that the decoupled version, $\decoup{\baseDist}$, of a base
distribution $\baseDist$ is obtained by combining two independent
samples of $\baseDist$, one for the first $t$ coordinates and the
other for the remaining.  A similar claim as above for the true acceptance
probabilities of the dictatorship test is now a simple corollary of
the above lemma and \cref{prop:bucketing_loss_bound}.  This will be
used later in extending the decoupling property to our general
inapproximability reduction.
}

\begin{lemma}
  \label{lemma:test soundness}
  For every predicate $\payoff$ and distribution $\baseDist$
  satisfying the conditions of \cref{theorem:reduction soundness},
  and any noise rate $\gamma > 0$, projection $\pi: \lcRightLabelSet \to \lcLeftLabelSet$, and
  bucketing parameter ${\buckParameter}$, the following holds.  For any functions $f:
  Q_1^\lcLeftLabelSet \to \Z$, $g: Q_2^\lcRightLabelSet \to \Z$ with bucketing functions $F:
  Q_1^\lcLeftLabelSet \to [{\buckParameter}]$, $G: Q_2^\lcRightLabelSet \to [{\buckParameter}]$,
  \begin{equation}\begin{split} \Big\lvert \accP(&\dicD) -  \accP(\dicDP) \Big\rvert\\%
      &\le \gamma^{-1/2} m^{1/2} 4^m {\buckParameter}^m%
      \sum_{a, b \in \irange{{\buckParameter}}} \Coinf{\super{F}{a},
        \super{G}{b}}^{1/2} + 2 {\buckParameter}^{-\delta} m^2.
\end{split}
\end{equation}
\end{lemma}

%


\def\finalDecouplingBound{ \gamma^{-1.5} m^{1/2} 4^m {\buckParameter}^{m+1} \val(\lc)^{1/2} + 2 {\buckParameter}^{-\delta} m^2 }

\label{sec:ocspInvariance}



\lncs{
  For space concerns, the proof of \cref{lemma:test soundness} is deferred to the full version.  Roughly, the idea is to prove a similar bound for the bucketed acceptance probability $\bVal(\dicD)$ and then use the \cref{prop:bucketing_loss_bound}.  The bound for the bucketed value goes via the invariance principle and uses a few sophisticated but standard estimates
  developed in the works of, amongst others, 
	Mossel~\cite{Mossel10}
and Wenner~\cite{Wenner12}. 
	With \cref{lemma:test soundness} in place, \cref{theorem:reduction soundness} can be derived using standard influence-decoding techniques; this is also deferred to the full version.
}

\full{
The proof of the theorem is via the invariance principle and uses a
few sophisticated but standard estimates developed in the works of
Mossel~\cite{Mossel10}, Samorodnitsky and
Trevisan~\cite{SamorodnitskyT09}, and Wenner~\cite{Wenner12}. 
The
first lemma essentially says that if a product of functions is
influential, then at least one of the involved functions is
influential. Applying the second lemma mostly involves introducing new
notation where the notion of \emph{lifted} functions is the most alien
-- a large-side table $g: \irange{q_2}^\lcRightLabelSet \rightarrow
\R$ may equivalently be seen as the function $\lifted[\pi]{g} :
\lifted[\lcLeftLabelSet]{\Omega} \rightarrow \R$ where
$\lifted{\Omega} = \irange{q_2}^d$ contains the values of all $d$
coordinates in $\lcRightLabelSet$ projecting to the same coordinate in
$\lcLeftLabelSet$. $\lifted[\pi]{g}$ is called the lifted analogue of
$g$ with respect to the projection $\pi$ and the remark below
essentially says that if the lifted analogue of $g$ is influential for
a coordinate $i \in \lcLeftLabelSet$, then $g$ is influential in a
coordinate projecting to $i$. The lemma will be used to -- after
massaging the expression -- decoupling the small-side table from the
lifted analogues of the large-side table as a function of their
\CoinfTerm{}.

\begin{lemma}[Lemma 6.5, Mossel~\cite{Mossel10}]
\label{lemma:influence_of_product_to_sum_of_influences}
Let $f_1, \dotsc, f_t: \Omega^n \rightarrow [0,1]$ be arbitrary. Then for any $j$,
$\Inf[j]{\prod_{r=1}^t f_r} \leq t \sum_{r=1}^t \Inf[j]{f_r}$.
\end{lemma}


\begin{remark}[Page 41, Wenner~\cite{Wenner12}]
\label{lemma:lifted_influences_to_projected}
Given a function $g: \Omega^\lcRightLabelSet \rightarrow
\lcRightLabelSet$ and a projection $\pi: \lcRightLabelSet \rightarrow
L$ where $\lcRightLabelSet = \lcLeftLabelSet \times \irange{d}$ and
$\lifted[\pi]{g}: (\Omega^d)^\lcLeftLabelSet \rightarrow \R$ is
suitably defined. Then the influence of a coordinate $i \in
\lcLeftLabelSet$ translates naturally to the sum of influences $j \in
\lcRightLabelSet$ projecting to $i$.
Namely, we have $ \Inf[i]{\lifted[\pi]{g}} = \Inf[\pi^{-1}(i)]{g} \leq
\sum_{j \varsuchthat \pi(j) = i} \Inf[j]{g}.  $ This follows from the
expression of influences in decompositions of $g$ which equals
$\sum_{T \varsuchthat i \in \pi(T)} \Ex{g_T^2}$ in the former two
cases and $\sum_T \card{T \cap \pi^{-1}(i)} \Ex{g_T^2}$ in the third.
\end{remark}

\begin{theorem}[Theorem 3.21, Wenner~\cite{Wenner12}]
\label{prop:yindepxinvariance}
\newcommand*{\stdprobspace}{\cP = (\prod_{i=1}^m \Omega_i, \rmP)}
Consider functions $\{f^{(r)} \in \rmL^\infty(\Omega_t^n)\}_{r \in \irange{m}}$
on a probability space $\stdprobspace^{\otimes n}$, a set $M \subsetneq \irange{m}$,
and a collection $\cC$ of minimal sets $C \subseteq [m], C \nsubseteq M$ such that the spaces $\{ \Omega_i \}_{i \in C}$ are dependent.
Then,
\begin{gather*}
  \abs{\Ex{\prod_{r=1}^m f^{(r)}} - \prod_{r \notin M} \Ex{f^{(r)}}
    \Ex{\prod_{r \in M} f^{(r)}}} \\\leq 4^m \max_{C \in \cC} \sqrt{
    \min_{r' \in C} \Totinf{f^{(r')}} \sum_l \prod_{r \in C \setminus
      \{r' \} } \Inf[l]{f^{(r)}} } \prod_{r \notin C}
  \norm[\infty]{f^{(r)}}.
\end{gather*}
\end{theorem}

\paragraph{Proof of \cref{lemma:test bucketed soundness}}
\begin{proof}
  We massage the expression $\bVal(\dicD)$ to a form suitable for applying
  \cref{prop:yindepxinvariance}.  Recall that $\bucketInd[a]$ denotes
  the indicator of ``$F(\rvx) = a$'' and similarly $\bucketIndR[a]$ of
  ``$G(\rvy) = a$''. Now, $\bVal(\dicD)$ equals
\[
\sum_{\vec{a} \in \irange{{\buckParameter}}^t, \vec{b} \in \irange{{\buckParameter}}^{m-t}}
\wp (\vec{a}, \vec{b})
\Ex[ \dicD ]{ \prod_{r=1}^t \super{F}{a_r}(\super{\rvvx}{r}) \prod_{r=t+1}^m \super{G}{b_r}(\super{\rvvy}{r}) },
\]
in terms of these indicators.  Consequently, $\abs{ \bVal(\mD) -
  \bVal(\decoup{\mD}) }$ may be bounded from above by
\begin{gather}
\label{eq:erite8ygioj}
\sum_{\vec{a}, \vec{b}}
\wp (\vec{a}, \vec{b})
\abs{
\Ex[ \ntestDist{\mD} ]{ \prod_{r=1}^t \super{F}{a_r}(\super{\rvvx}{r}) \prod_{r=t+1}^m \super{G}{b_r}(\super{\rvvy}{r}) }
-
\Ex[ \ntestDist{\decoup\mD} ]{ \prod_{r=1}^t \super{F}{a_r}(\super{\rvvx}{r}) \prod_{r=t+1}^m \super{G}{b_r}(\super{\rvvy}{r}) }
}.
\end{gather}%

We note that $\wp$ is bounded by $1$ in magnitude and proceed to bound
the expression inside the summation.  To this end, we must make a
slight change of notation as discussed previously. The new notation
may seem cumbersome; the high-level picture is that we group the first
set of functions into a single function and redefine the latter to be
functions on arguments indexed by $\lcLeftLabelSet$ instead of
$\lcRightLabelSet$.

Define $m' = m - t + 1$, $\Omega_1 = \irange{q_1}^t$, $\Omega_2 =
\dotsc = \Omega_{m'} = \irange{q_2}^d$ and let $\pcomp{\gamma}$ denote
$1 - \gamma$. Let $\varsigma$ be a bijection $\lcLeftLabelSet \times
\irange{d} \leftrightarrow \lcRightLabelSet$ such that
$\pi(\varsigma(i, j')) = i$. Introduce the distribution
$\Omega_1^\lcLeftLabelSet \times \dotsc \Omega_{m'}^\lcLeftLabelSet
\ni (\rvv{w}, \super{\rvvz}{2}, \dotsc, \super{\rvvz}{m'}) \sim
\cR(\mu)$ which samples $(\super{\rvvx}{1}, \dotsc, \super{\rvvx}{t},
\super{\rvvy}{t+1}, \dotsc, \super{\rvvy}{m})$ from $\dicD$, setting
$\rv{w}_{i,r} = \super{\rvx}{r}_i$ and $\super{\rvz}{r}_i = \{
\super{\rvy}{r}_{\varsigma^{-1}(i,j')} \}_{j'=1}^d$. Let $W(\vec{w})
\defeq \prod_{r=1}^t
(\noiseop_{\pcomp{\gamma}}\super{F}{r})(\super{\vx}{r})$ where
$\super{x}{r}_i = w_{i,r}$ and similarly, for $2 \leq t \leq m'$, call
the lifted function $\super{H}{r-t+1}: \Omega_t^L \rightarrow \R$
defined as $\super{H}{r-t+1}(\vz) =
(\noiseop_{\pcomp{\gamma}}\super{G}{r})(\vy)$ where
$y_{\varsigma(i,j')} = z_{i,j'}$. With this new notation, the
difference within the summation in (\ref{eq:erite8ygioj}) is
\begin{gather}
\label{eq:dfsjkddopu8390}
\abs{
\Ex[\cR(\mD)]{ W(\rvv{w}) \prod_{r=2}^{m'} \super{H}{r}(\super{\rvvz}{r}) }
- \Ex[\cR(\decoup{\mD})]{ W(\rvv{w}) \prod_{r=2}^{m'} \super{H}{r}(\super{\rvvz}{r}) }
}.
\end{gather}

We note that $\cR$ is a product distribution $\cR = \mu^{\otimes \lcLeftLabelSet}$ for some $\mu$ and for any $2 \leq t \leq m'$, $\super{\Omega}{1}$ is independent of $\super{\Omega}{t}$. Choosing $M = \{2, \dotsc, m'\}$, minimal indices $\cC$ of dependent sets in $\mu$ not contained in $M$ contains 1 and at least two elements from $M$, i.e.\ $C \in \cC$ implies $1,e,e' \in C$ for some $e \neq e' \in \{2,\dotsc,m'\}$.

Applying \cref{prop:yindepxinvariance} and choosing $r' \neq 1$ bounds the difference (\ref{eq:dfsjkddopu8390}) by
\begin{gather}
\label{eq:i35r90weyfiosjdfiop}
4^m \sqrt{ \max_{C \in \cC} \max_{e \in C} \Totinf(\super{H}{e}) \sum_i \Inf[i]{W} \prod_{t \in C \setminus \{1, e \} } \Inf[i]{\super{H}{t}} }
\prod_{r \notin C} \norm[\infty]{\super{H}{r}}.
\end{gather}

As we assumed the codomain of the studied functions $\{\super{G}{r}\}_r$ to be $[0,1]$ the same holds for $\{\super{H}{r}\}_r$ and consequently the influences and infinity norms in (\ref{eq:i35r90weyfiosjdfiop}) are upper-bounded by one on account of \cref{lemma:influence_variance_bound}, yielding
\begin{gather}
\label{eq:34950wtuefsdo}
(\ref{eq:dfsjkddopu8390}) \leq 4^m \left( \max_e \Totinf (\super{H}{e}) \cdot \max_e \sum_i \Inf[i]{W} \Inf[i]{\super{H}{e}} \right)^{1/2}.
\end{gather}

We recall that $W = \prod_{t=1}^r \noiseop_{\pcomp{\gamma}} \super{f}{r}$ and hence by \cref{lemma:influence_of_product_to_sum_of_influences}, $\Inf[i]{W} \leq t \sum_{r=1}^t \Inf[i]{\noiseop_{\pcomp{\gamma}} \super{f}{r}} = t \sum \Inf[i][\pcomp{\gamma}]{\super{f}{r}}$.
Similarly, \cref{lemma:lifted_influences_to_projected} implies that $\Inf[i]{\super{H}{e-t+1}} \leq \sum_{j \in \pi^{-1}} \Inf[j]{\noiseop_{\pcomp{\gamma}}\super{G}{e}} = \Inf[j][\pcomp{\gamma}]{ \super{G}{e}}$.
 Returning to (\ref{eq:34950wtuefsdo}), we have the bound
\begin{gather*}
(\ref{eq:dfsjkddopu8390}) \leq 4^m \left( t \max_e \Totinf[\pcomp{\gamma}](\super{G}{e}) \max_e \sum_i \sum_{r=1}^t \Inf[i][\pcomp{\gamma}]{\super{f}{r}} \sum_{j \in \pi^{-1}(i)} \Inf[j][\pcomp{\gamma}]{\super{G}{e}} \right)^{1/2}.
\end{gather*}
Bounding the total influence using \cref{lemma:total_noisy_influence_bound} and identifying the inner sum as a \CoinfTerm, we establish the desired bound on the lemma difference
\begin{gather*}
(\ref{eq:dfsjkddopu8390}) \leq 4^m \left( t \gamma^{-1} \sum_{r,e} \Coinf{\super{f}{r}, \super{G}{e}} \right)^{1/2}
\leq \gamma^{-1/2} m^{1/2} 4^m \sum_{r,r'} \Coinf{\super{f}{r}, \super{G}{r'}}^{1/2}.
\end{gather*}
Finally, as we noted before, since $0 < \wp < 1$,
(\ref{eq:erite8ygioj}) is at most
\begin{align*}
\gamma^{-1/2} m^{1/2} 4^m {\buckParameter}^m \sum_{a, b \in \irange{{\buckParameter}}} \Coinf{\super{F}{a}, \super{G}{b}}^{1/2},
\end{align*} as there are at most $\Gamma^m$ terms in the summation.
\qedendproof
}




\subsection{Soundness of the Reduction}
\label{sec:reduction soundness}

With the soundness for the dictatorship test in place, proving the
soundness of the reduction (\cref{theorem:reduction soundness}) is a
relatively standard task of 
%
%
constructing noisy-influence decoding
strategies. 


The proof follows immediately from the a more general estimate given
in the following Lemma by taking ${\buckParameter} = (4m^2/\epsilon)^{1/\delta}$ and
then $\epsilon_{LC} = \left(\frac{\epsilon \gamma^{3/2}}{m^{1/2} 4^m
    {\buckParameter}^{m+1}}\right)^2$.
\newcommand{\strat}[1][{}]{\Psi_{#1}}%
\begin{lemma}\label{prop:expBound} Given an \labelcover instance $\lcDesc$
  and a collection of functions, $f_u: Q_1^L \to \Z$ for $u \in
  \lcLeft$; $g_v: Q_2^R \to \Z$ for $v \in \lcRight$, and ${\buckParameter}$,
  $\gamma$, $\delta$ as in this section,
    \[
    \Ex[u,v \sim \lcEdges]{ \left| \accP[f_u, g_v](\dicD) - \accP[f_u, g_v](\dicDP) \right|} \le
    \finalDecouplingBound.
    \]
\end{lemma}

\begin{proof}
  For a function $f: Q_1^L \to \Z$  define
  a distribution $\strat(f)$ over $L$ as follows.  First pick $a \sim \irange{{\buckParameter}}$ uniformly, then pick $l \in L$ with probability $\gamma \cdot \infl(\bucketInd[a]_v)$ and otherwise an arbitrary label.  
  Note that by \cref{prop:totInfBound}, $\sum_{l \in L} \infl(\bucketInd[a]_u) \le 1/\gamma$ and so picking $l \in L$ with the given probabilities is possible.  Define $\strat(g)$ over
  $R$ for $g: Q_2^R \to \Z$ similarly.  
  Now define a labeling of $\mathcal{L}$ by, for each $u \in U$
  (resp.~$v \in V$), sampling a label from $\strat(f_u)$
  (resp.~$\strat(g_v)$) independently.
  
  For an edge $e = (u,v) \in \lcEdges$, the probability that $e$ is satisfied by the labeling equals $\prob{ \pi_e(\strat(f_u)) = \strat(g_v)}$, which can be lower bounded by
  \begin{gather*}
    \sum_{i,j \varsuchthat \pi_e(j) = i} \gamma^2
    \Ex[a,b \in \irange{{\buckParameter}}]{ \Inf[i][1-\gamma]{\bucketInd[a]_u} \Inf[j][1-\gamma]{\bucketIndR[b]_v} }
    = (\gamma/{\buckParameter})^2
    \sum_{a,b} \Coinf[\pi_e]{\bucketInd[a]_u, \bucketIndR[b]_v }.
  \end{gather*}
  Taking the expectation over all edges of $\mathcal{L}$ we get that the fraction of satisfied constraints is
  \[
  (\gamma/{\buckParameter})^2 \E_{e = (u,v)}\left[\sum_{a,b} \Coinf[\pi_e]{\bucketInd[a]_u,
    \bucketIndR[b]_v }\right] \le \val(\mathcal{L}),
\] and by concavity of the $\sqrt{\cdot}$ function, this implies that
$\E_{e = (u,v)}\left[\sum_{a,b}
  \Coinf[\pi_e]{\bucketInd[a]_u,\bucketIndR[b]_v }^{1/2}\right] \le
{\buckParameter}\gamma^{-1} \val(\mathcal{L})^{1/2}$.
  Plugging this bound on the total cross influence into the soundness
  for the dictatorship test \cref{lemma:test soundness}, we obtain
  \[
  \Ex[e = (u,v)]{ \left| \accP[f_u, g_v](\dicDpie) - \accP[f_u, g_v](\dicDPpie) \right|} \le
  \finalDecouplingBound,
  \]
  as desired.
\qedendproof


%% file: conclusion.tex
\section*{Conclusion}

We gave improved inapproximability for several important OCSPs.  Our
characterization is by no means complete and several interesting
problems are still open.  Closing the gap in the approximability of
\MAS is wide open and probably no easier than resolving the
approximability for \textproblem{Max Cut} and other $2$-CSPs.  In
particular, getting any factor close to $1/2$ seems to require new
ideas.  \MaxBTW has an approximation algorithm that satisfies half of
the constraints if all the constraints can be simultaneously
satisfied~\cite{ChorS98}.  Thus improving our result to obtaining
perfect completeness is particularly enticing.  Finally, though the
inability to fold long codes is a serious impedement, improving our
general hardness result to only requiring that $\baseDist$ is pairwise
independent is interesting especially in light of the analogous
results for CSPs~\cite{AustrinM09,Chan13}.

{
\small
\paragraph{Acknowledgement.} We would like to thank \JHastad for
suggesting this problem and for numerous helpful discussions regarding
the same.  We acknowledge ERC Advanced Grant $226203$ and Swedish
Research Council Grant $621$-$2012$-$4546$ for making this project
feasible.  }
